\documentclass[amssymb,amsmath,prd,twocolumn,superscriptaddress,nofootinbib]{revtex4-1}

\usepackage[10pt]{type1ec}  
\usepackage[T1]{fontenc}
\usepackage{CJKutf8}
\usepackage[overlap, CJK]{ruby}
\usepackage{CJKulem}
\newenvironment{SChinese}{%
  \CJKfamily{gbsn}%
  \CJKtilde
  \CJKnospace}{}

\usepackage{graphicx}
\usepackage{bm}
\usepackage{mathrsfs}
\usepackage{latexsym}
\usepackage{color}
\usepackage{dcolumn}
\usepackage[colorlinks=true,citecolor=blue,urlcolor=blue]{hyperref}
\usepackage[usenames,dvipsnames]{xcolor}
\usepackage[caption=false]{subfig}

\newcommand{\p}{\partial}

\newcommand{\beq}{\begin{equation}}
\newcommand{\eeq}{\end{equation}}
\newcommand{\ba}{\begin{array}}
\newcommand{\ea}{\end{array}}
\newcommand{\bea}{\begin{eqnarray}}
\newcommand{\eea}{\end{eqnarray}}
\newcommand{\bc}{\begin{center}}
\newcommand{\ec}{\end{center}}

\newcommand{\be}{\beta}

\newcommand{\De}{\Delta}
\newcommand{\ep}{\varepsilon}

\newcommand{\La}{\Lambda}

\newcommand{\ptrans}{p_{\rm trans}}
\newcommand{\ntrans}{n_{\rm trans}}
\newcommand{\nb}{n_{\rm B}}
\newcommand{\etrans}{\varepsilon_{\rm trans}}
\newcommand{\mutrans}{\mu_{\rm trans}}
\newcommand{\Mtrans}{M_{\rm trans}}

\newcommand{\Msolar}{{\rm M}_{\odot}}

\newcommand{\ncent}{n_{\rm cent}}
\newcommand{\mucent}{\mu_{\rm cent}}

\newcommand{\cQMsq}{c^2_{\rm QM}}
\newcommand{\csqtwo}{c^2_{s, 2.0}}
\newcommand{\csqtyp}{c^2_{s, 1.4}}
\newcommand{\cNMsq}{c^2_{\rm NM}}

\newcommand{\Mmax}{M_{\rm max}}
\newcommand{\Rmax}{R_{\rm max}}
\newcommand{\nmax}{\ncent^{\rm max}}
\newcommand{\csqmax}{c^2_{s, {\rm max}}}
\newcommand{\mumax}{\mucent^{\rm max}}
\newcommand{\Rtyp}{R_{1.4}}
\newcommand{\Rtwo}{R_{2.0}}

\newcommand{\Mchirp}{{\mathcal M}}

\newcommand\Eqn[1]{Eq.~(\ref{#1})}  

\allowdisplaybreaks

\begin{document}

\title{On the minimum radius of very massive neutron stars}

\author{Sophia Han 
(\begin{CJK}{UTF8}{}\begin{SChinese}韩 君\end{SChinese}\end{CJK})
}
\email{sjhan@berkeley.edu}
\affiliation{Department of Physics and Astronomy, Ohio University,
Athens, OH~45701, USA}
\affiliation{Department of Physics, University of California Berkeley, Berkeley, CA~94720, USA}

\author{Madappa Prakash} 
\email{prakash@ohio.edu} 
\affiliation{Department of Physics and Astronomy, Ohio University,
Athens, OH~45701, USA}

\date{June 28, 2020}

\begin{abstract}

Prospects of establishing the radii of massive neutron stars in PSR J1614-2230 and PSR J0740+6620 from {\it{NICER}} and {\it{Chandra}} observatories hold the potential to constrain the equation of state (EoS) of matter to densities well beyond those encountered in canonical stars of mass $\sim 1.4\,\Msolar$. In this work, we investigate the relation between the radii of very massive neutron stars up to the maximum mass, $\Mmax$, supported by dense matter EoSs. Results from models with hadronic matter are contrasted with those that include a first-order hadron-to-quark phase transition. We find that a lower bound on $\Mmax$ with an upper bound on the radius of massive pulsars serves to rule out too soft quark matter, and an upper bound on $\Mmax$ with a lower bound on the radius of massive pulsars strongly disfavors a transition into too-stiff quark matter appearing at low densities. The complementary role played by radius inferences from future gravitational wave events of inspiraling binary neutron stars is also briefly discussed.

\end{abstract}

\maketitle
  
\section{Introduction}
\label{sec:intro}

The two most basic properties of a neutron star are its mass $M$ and radius $R$. The importance of these physical traits is highlighted by their influence on several other neutron star (NS) observables that include 
(i) the binding energy, $\rm{B.E.}$ $\propto \be M$, where $\be=GM/Rc^2$ is the compactness, (ii) spin periods of rotation, $P \propto M^{-1/2} R^{3/2}$, (iii) moment of inertia, $I \propto MR^2$, (iv) tidal deformability, $\La \propto \be^{-6}$, etc. For a list of other observables that are significantly influenced by $M$ and $R$, see Ref.~\cite{Lattimer:2015nhk}. Through general relativistic equations of the stellar structure \cite{Tolman:1939jz,Oppenheimer:1939ne}, rotation~\cite{Friedman:1986tx}, and tidal deformations~\cite{Damour:2009vw,Hinderer:2009ca,Postnikov:2010yn}, these observables can be calculated once the relationship between pressure $p$ and energy density $\ep$, or the equation of state (EoS), of neutron star matter is provided. The one-to-one-correspondence between the EoS ($p$ vs. $\ep$) and the observed $M$-$R$ curve can then be used to advantage to model-independently determine the EoS of neutron star matter~\cite{Lindblom:1992}.

The number of neutron stars for which simultaneous measurements of masses and radii along with inherent systematic and statistical errors are available will determine the accuracy with which the EoS can be determined. To date, data on radii to the same level of accuracy that radio pulsar measurements on masses of neutron stars have afforded us do not exist. 
The discovery of neutron star masses up to and beyond $2\,\Msolar$~\cite{Demorest:2010bx,Fonseca:2016tux,Arzoumanian:2017puf,Antoniadis:2013pzd,Cromartie:2019kug,Linares:2018ppq} has not only spurred a great deal of theoretical activity, but emphasized the need to measure the radii of such stars so that the physics of strong interactions at high baryon density can be better understood.

\begin{table*}
\begin{center}
\begin{tabular}{l@{\quad}c@{\quad}c@{\quad}c@{\quad}@{\quad}c}
\hline \hline
&&&\\[-2ex]
Source & Radius (km) & Mass ($\Msolar$) & References \\[0.5ex]
\hline
&&&\\[-2ex]
X-ray Observations & $9-14$ & $\sim 1.4$ & \cite{Lattimer:2012nd,Ozel:2016oaf} \\[0.5ex]
& $10-14$ & $\sim1.4$ & \cite{Steiner:2015aea,Steiner:2017vmg} \\[0.5ex]
\hline
&&&\\[-2ex]
GW170817 & $8.9-13.2$ & ${1.36\, (1.17)-1.60\, (1.36)}$ & \cite{De:2018uhw} \\[0.5ex]
& $11\pm 1$ & ${1.36\, (1.17)-1.60\, (1.36)}$ & \cite{Capano:2019eae} \\[0.5ex]
\hline
&&&\\[-2ex]
{\it NICER} & $13.02^{+1.24}_{-1.19}$ & $1.44^{+0.15}_{-0.14}$ & \cite{Miller:2019cac} \\[0.5ex]
(PSR J0030+0451) & $12.71^{+1.14}_{-1.19}$ & $1.34^{+0.15}_{-0.16}$ & \cite{Riley:2019yda} \\[0.5ex]
\hline
\end{tabular}
\end{center}
\caption{Estimates of radii and masses of neutron stars.}
\label{tab:radii}
\end{table*}

Table~\ref{tab:radii} provides a summary of available radius estimates from x-ray observations of neutron stars (reviewed in Refs.~\cite{Lattimer:2012nd,Ozel:2016oaf}) including the recent results~\cite{Riley:2019yda,Miller:2019cac} from the {\it {Neutron Star Interior Composition ExploreR (NICER)}} and from the binary neutron star merger event GW170817~\cite{LIGO:2017qsa,De:2018uhw,Capano:2019eae}. 
Owing to large systematic errors, analyses of x-ray data yield radii in the range of $\approx (9$-$14$) km for neutron star masses in the vicinity of the canonical $1.4\,\Msolar$~\cite{Bogdanov:2012md,Guillot:2013wu,Steiner:2015aea,Steiner:2017vmg,Nattila:2015jra,Nattila:2017wtj,Ozel:2015fia}. 
Although analysis of the data during inspiral (the phase prior to coalescence) by the LIGO and Virgo collaborations has yielded the chirp mass $\Mchirp=(m_{1}m_{2})^{3/5}/(m_1+m_2)^{1/5}=1.186_{-0.001}^{+0.001}\Msolar$~\cite{Abbott:2018wiz}, where $m_1$ and $m_2$ are the component masses and the total mass $m_{\rm tot}=m_1+m_2=2.73^{+0.04}_{-0.01}\,\Msolar$, constraints on the masses and radii inferred through the combined tidal deformabilities are not very restrictive (see Table~\ref{tab:radii} and the references therein). 
Recent results from {\it{NICER}} fall in a different category insofar as its aim has been to measure the mass and radius of the same neutron star. It is heartening that two independent studies~\cite{Riley:2019yda,Miller:2019cac} have reported results that are consistent with each other within the quoted statistical and systematic errors. It is noteworthy that the neutron star masses involved in all of these studies hover around the canonical mass of $\sim 1.4\,\Msolar$. 

As noted earlier, the determination of the dense matter EoS depends on the simultaneous knowledge of masses and radii of individual neutron stars over a wide range of masses, including those close to the as yet unknown maximum mass. 
Combining the electromagnetic (EM) and gravitational wave (GW) information~\cite{LIGO:2017qsa,GBM:2017lvd} from GW170817, Ref.~\cite{Margalit:2017dij} has limited the maximum gravitational mass $M^g_{\rm max}$ and radius of a neutron star as $M^g_{\rm max} \lesssim 2.17\, \Msolar$ and $R_{1.3} \gtrsim 3.1\,GM^g_{\rm max} \simeq 9.92~{\rm km} $, where $R_{1.3}$ is the radius of a $1.3\,\Msolar$ neutron star. 
These estimates have been revisited in Ref.~\cite{Shibata:2019ctb} where a weaker constraint $M^g_{\rm max} \lesssim  2.3 \,\Msolar$ has been reported. 
Utilizing the total mass of $2.74^{+0.04}_{-0.01}\,\Msolar$ from GW170817 with (empirical) universal relations between the baryonic and the maximum rotating and non-rotating masses of neutron stars, Ref.~\cite{Rezzolla:2017aly} constrained the maximum non-rotating neutrons star mass in the range $2.01^{+0.04}_{-0.01}\,\Msolar - 2.16^{+0.17}_{-0.15}\,\Msolar$. Ref.~\cite{Ruiz:2017due} combined general relativistic magnetohydrodynamic simulation results and the discovery of GW170817, yielding the maximum mass of a cold, spherical neutron star in the range $2.16 - 2.28 \,\Msolar$. 
Interestingly, these limits are close to the well measured masses of the very massive neutron stars listed in Table \ref{tab:masses}. Observational determination of the radii of these massive stars would greatly advance theoretical efforts to delimit the EoS of dense matter.

\begin{table}[]
\begin{center}
\begin{tabular}{l@{\quad}c@{\quad}c@{\quad}c}
\hline \hline
&&&\\[-2ex]
Source & Mass $(\Msolar)$ & References   \\[0.5ex]
\hline  
&&&\\[-2ex]
PSR J1614-2230& $1.97\pm 0.04$ &\cite{Demorest:2010bx}   \\[0.5ex]
&&&\\[-2ex]
 & $1.928 \pm 0.017$ & \cite{Fonseca:2016tux}  \\[0.5ex]
&&&\\[-2ex]
 & $1.908 \pm 0.016$ & \cite{Arzoumanian:2017puf}  \\[0.5ex]
&&&\\[-2ex]
PSR J0348+0432 & $2.01 \pm 0.04$  & \cite{Antoniadis:2013pzd}    \\[0.5ex]
&&&\\[-2ex]
PSR J0740+6620 & $2.14^{+0.10}_{-0.09} $ & \cite{Cromartie:2019kug}   \\[0.5ex] 
&&&\\[-2ex]
PSR 2215-5135 & $2.27^{+0.17}_{-0.15}$ & \cite{Linares:2018ppq}      \\[0.5ex]
\hline
\end{tabular}
\end{center}
\caption{Largest measured masses of neutron stars.}
\label{tab:masses}
\end{table}  

A proposal to observe PSR J1614-2230 with its mass $\simeq 1.93\,\Msolar$~\cite{Fonseca:2016tux} (later updated to $\simeq 1.91\,\Msolar$~\cite{Arzoumanian:2017puf}) utilizing {\it{NICER}} in concert with {\it{Chandra}} observations (to establish the background) has been put forth by Ref.~\cite{Miller:2016kae}. 
Although upper limits on the radius will be difficult given the expected low count rate, lower limits on the radius could be established. 
Upcoming analyses of {\it{NICER}} data on PSR J1614-2230 as well as PSR J0740+6620 with its mass $\simeq2.14 \,\Msolar$ \cite{Cromartie:2019kug} will for the first time provide constraints on the unknown radii of these very massive neutron stars.

\begin{figure*}[htb]
\parbox{0.35\hsize}{
\includegraphics[width=\hsize]{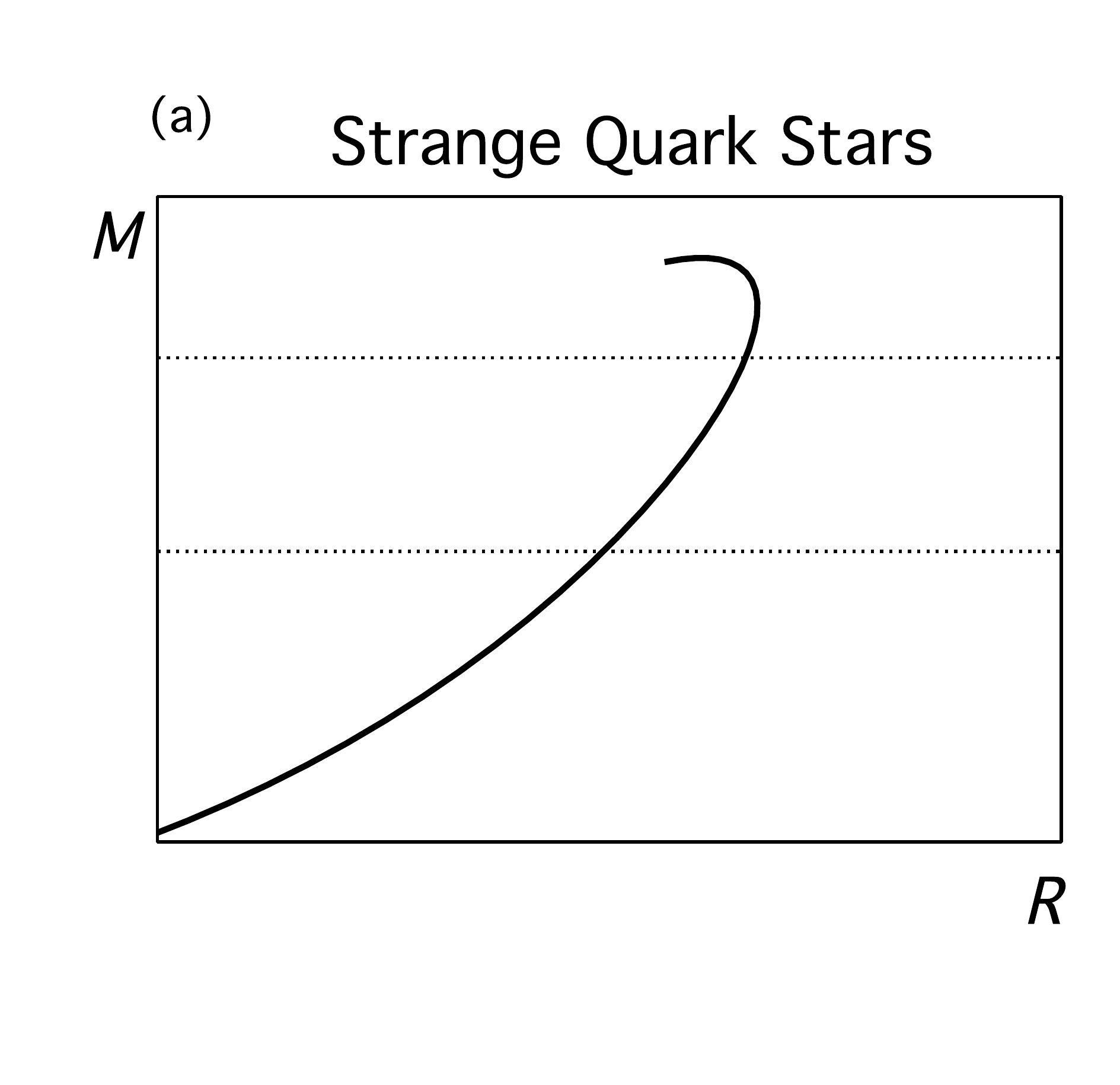}\\[-2ex]
}\parbox{0.35\hsize}{
\includegraphics[width=\hsize]{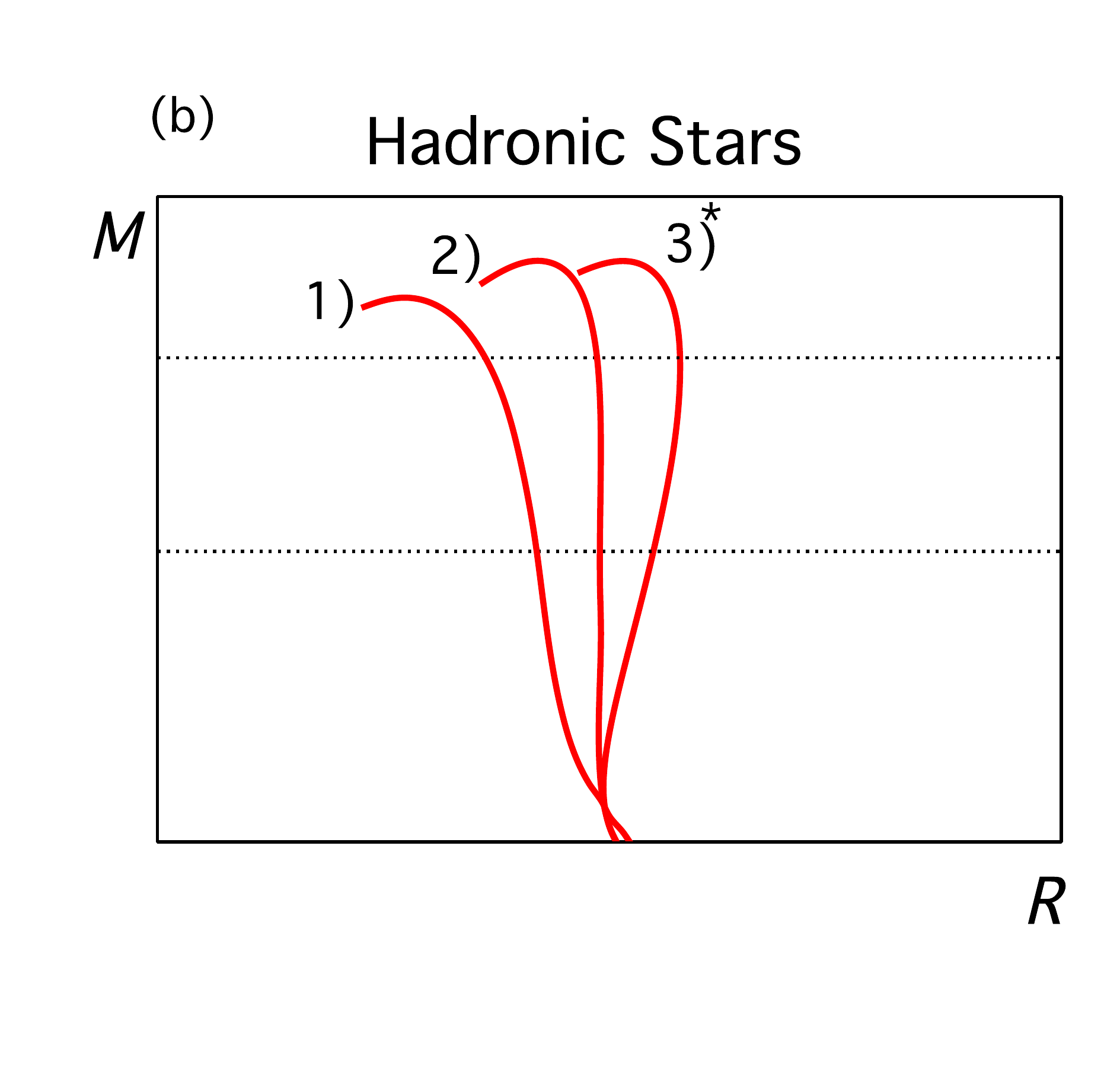}\\[-2ex]
}\parbox{0.35\hsize}{
\includegraphics[width=\hsize]{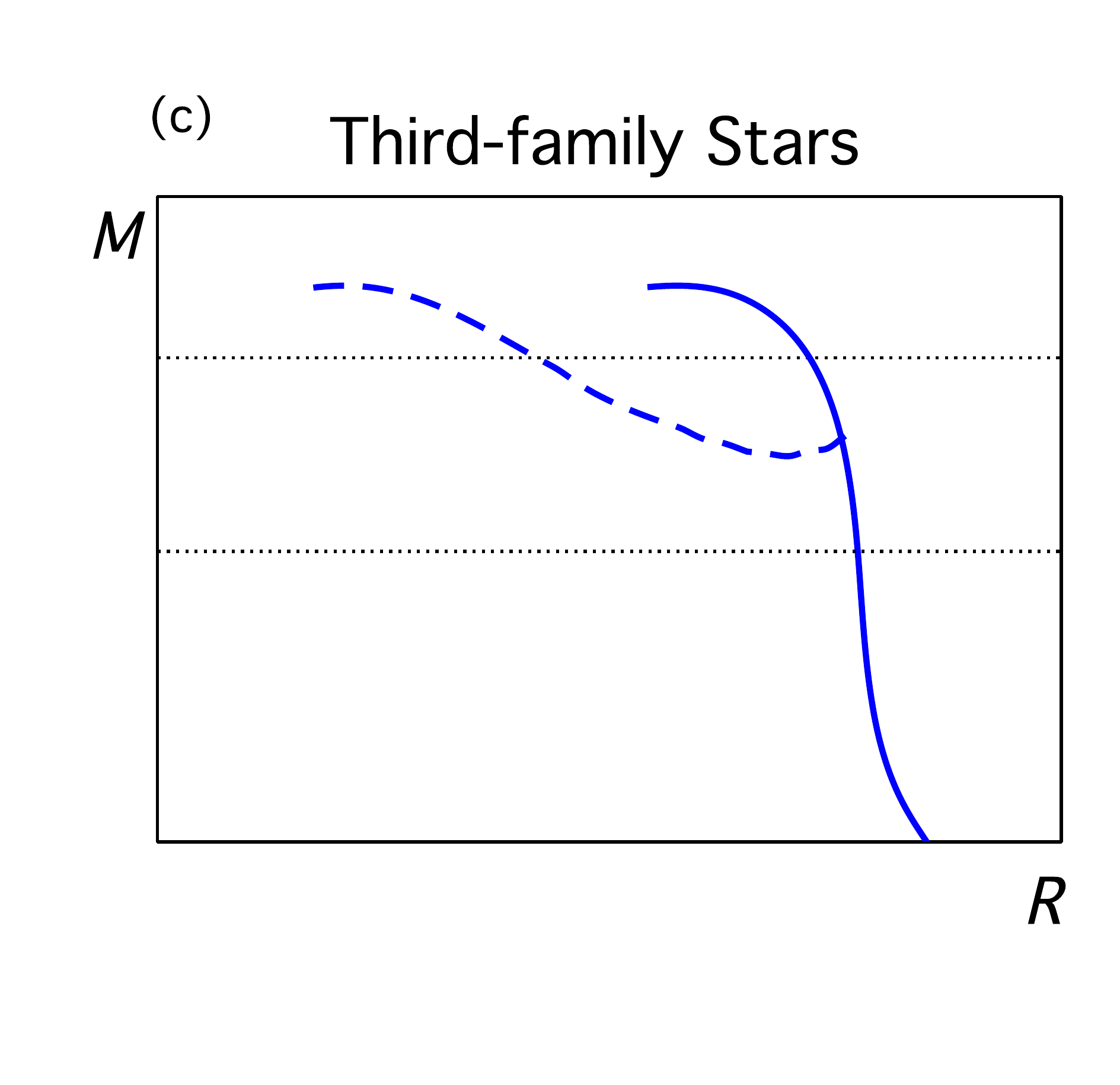}\\[-2ex]
}
\caption{Schematic diagram showing representative mass-radius relations for self-bound strange quark stars (panel (a)), normal hadronic stars and those with a smooth phase transition (panel (b)), and hybrid stars with a sufficiently strong first-order transition that form a ``third family'' (panel (c)). The two horizontal dotted lines indicate $M=2.0\,\Msolar$ and $M=1.4\,\Msolar$. For self-bound stars, $\Rtwo>\Rtyp$ is always obeyed. For hadronic models, the relation between $\Rtyp$ and $\Rtwo$ varies: (1) $\Rtwo<\Rtyp$, (2) $\Rtwo\approx\Rtyp$, and $(3)^{*}$ $\Rtwo>\Rtyp$. The two hadronic EoSs used in the present paper (HLPS and DBHF) both belong to case (1). Note that in case $(3)^{*}$, $\Rtwo>\Rtyp$ is also possible in quarkyonic models for which the EoS is continuous, but exhibits a sudden stiffening around $2\sim2.5 \, n_{0}$ (see e.g. Fig. 8 in Ref.~\cite{Zhao:2020dvu}), and in hybrid EoSs with a sharp first-order transition at low densities (see e.g. Appendix~\ref{AppB}). ``Third-family stars'' are a distinctive subset of hybrid EoSs that are maximally distinguishable from those in panel (b), but not necessarily realized in nature (see e.g. Appendix~\ref{AppB}). 
}
\label{fig:MR-illust}
\end{figure*}

In view of the above prospect, it is the purpose of this paper to explore the minimum radius of very massive stars based on theoretical models of neutron star structure. In addition to models with hadronic matter, we also consider those in which a first-order hadron-to-quark phase transition takes place. 
An added impetus for this study is the recent discovery of the binary merger GW190425 with a total mass of $\sim 3.4\,\Msolar$~\cite{Abbott:2020uma}, which points to at least one of the companion's mass to be significantly more massive than in GW170817. 
The categories of the mass-radius ($M$-$R$) relations of neutron stars studied in this work are discussed in Sec.~\ref{sec:class}. 
In Sec.~\ref{sec:rmax}, $R_{2.0}$, radii of $2\,\Msolar$ stars for models with and without hadron-to-quark phase transition are contrasted. This section also provides a discussion of the sensitivity of results to the near- and supra-nuclear behaviors of the EoS models considered. 
Sec.~\ref{sec:obs} highlights how constraints on radii of neutron stars more massive than the canonical $1.4\,\Msolar$ value from future {\it{NICER}} and GW measurements play complementary roles. 
Our conclusions are contained in Sec.~\ref{sec:con}.

\section{Categories of $M$-$R$ relations}
\label{sec:class}

Figure~\ref{fig:MR-illust} illustrates the landscape of $M$-$R$ curves resulting from the use of various possible EoSs. The examples shown in this figure have been chosen to meet the criterion of satisfying the $2\,\Msolar$ constraint required by observations of massive stars. The three categories shown correspond to distinct classes of EoSs. 
Those in panels (b) and (c) tally with laboratory data on nuclei near their equilibrium density of $n_0\simeq 0.16\pm 0.01~{\rm fm^{-3}}$, but differ in their physical descriptions at higher densities (see below). The characteristic feature of the behavior shown in panel (c) (termed ``third-family stars'') is that an unstable branch is encountered prior to a second stable branch in which $M \geq 2\,\Msolar$. 
The $M$-$R$ curve in panel (a) is that of a self-bound strange quark star~\cite{Witten:1984rs,Farhi:1984qu}. 
Distinguishing traits of these three cases are briefly described below.

Self-bound strange quark stars (SQSs) have a sharp surface between high-density quark matter and the vacuum, where the pressure abruptly vanishes below a few times $n_0$. They exhibit a distinctive $M$-$R$ relation in that the radius increases with the mass till approaching the limit of collapsing into a black hole. Note that for a SQS, $\Rtyp<\Rtwo$ is always satisfied. It is generally accepted that observations of thermal emissions from a neutron star surface and data on pulsar glitches favor a standard crust formation. Although SQSs form a separate family of neutron stars, observational evidence of their existence has remained elusive.

Little-to-moderate variation in the radius is seen for normal hadronic stars in the mass range $1.1-1.7\,\Msolar$. This is because radii of such stars are mostly sensitive to the nuclear symmetry energy around $1-3\,n_0$~\cite{Lattimer:2000nx}. For masses reaching up to the maximum mass, a decrease in the corresponding radii of $\lesssim 1.5$~km is not uncommon.
Modern calculation of neutron matter, from e.g., chiral effective field theory with error estimates, are regarded reliable only for $\nb\lesssim 1.5-2\,n_{0}$ owing to the perturbative expansion scales reaching invalid regions as the density increases toward the central densities of maximum mass stars.
To construct a full EoS for neutron-star matter from this approach, calculations have relied on non-trivial extrapolations from $\sim1.5\,n_{0}$ up to the central density of the maximum-mass NS $\nb \gtrsim 4-8 \,n_{0}$. Assuming specific models or applying parametrizations consistent with causality, consistency with observational constraints have been achieved with a rapid stiffening in the EoS at $\sim 2\,n_0$ strongly indicated. 
It is worthwhile pointing out that many phenomenological descriptions based on non-relativistic potential models with causality enforced and relativistic mean-field theoretical models with suitable extensions have also succeeded in satisfying the current astrophysical constraints (see e.g. Ref.~\cite{Oertel:2016bki} for a review). 
For our purposes in this work, it suffices to note the three possibilities $\Rtyp \gtreqqless \Rtwo$ that this class of models yield.

At the dense cores of neutron stars, exotic particles (hyperons, kaon/pion condensates or quarks) may emerge and even become abundant. Hybrid stars with transitions from nucleonic matter into exotic matter have been extensively studied using various treatments, see e.g., Ref.~\cite{Han:2019bub}.
These include a first-order phase transition with a sharp interface (Maxwell construction)~\cite{Glendenning:1992vb} or through a mixed phase (Gibbs construction)~\cite{Glendenning:2001pe}, a smooth hadron-quark crossover with an interpolation scheme \cite{Baym:2017whm} or as in the quarkyonic matter model~\cite{McLerran:2018hbz}. 
It turns out that for hybrid EoSs with weak or smooth transitions, the masquerade problem~\cite{Alford:2004pf} prevents one from distinguishing them from normal hadronic stars. However, if there is a strong enough first-order transition between phases of vastly different density, then this could verifiably affect the static properties such as mass, radius, and tidal deformability.

\begin{figure*}[htb]
\parbox{0.33\hsize}{
\includegraphics[width=\hsize]{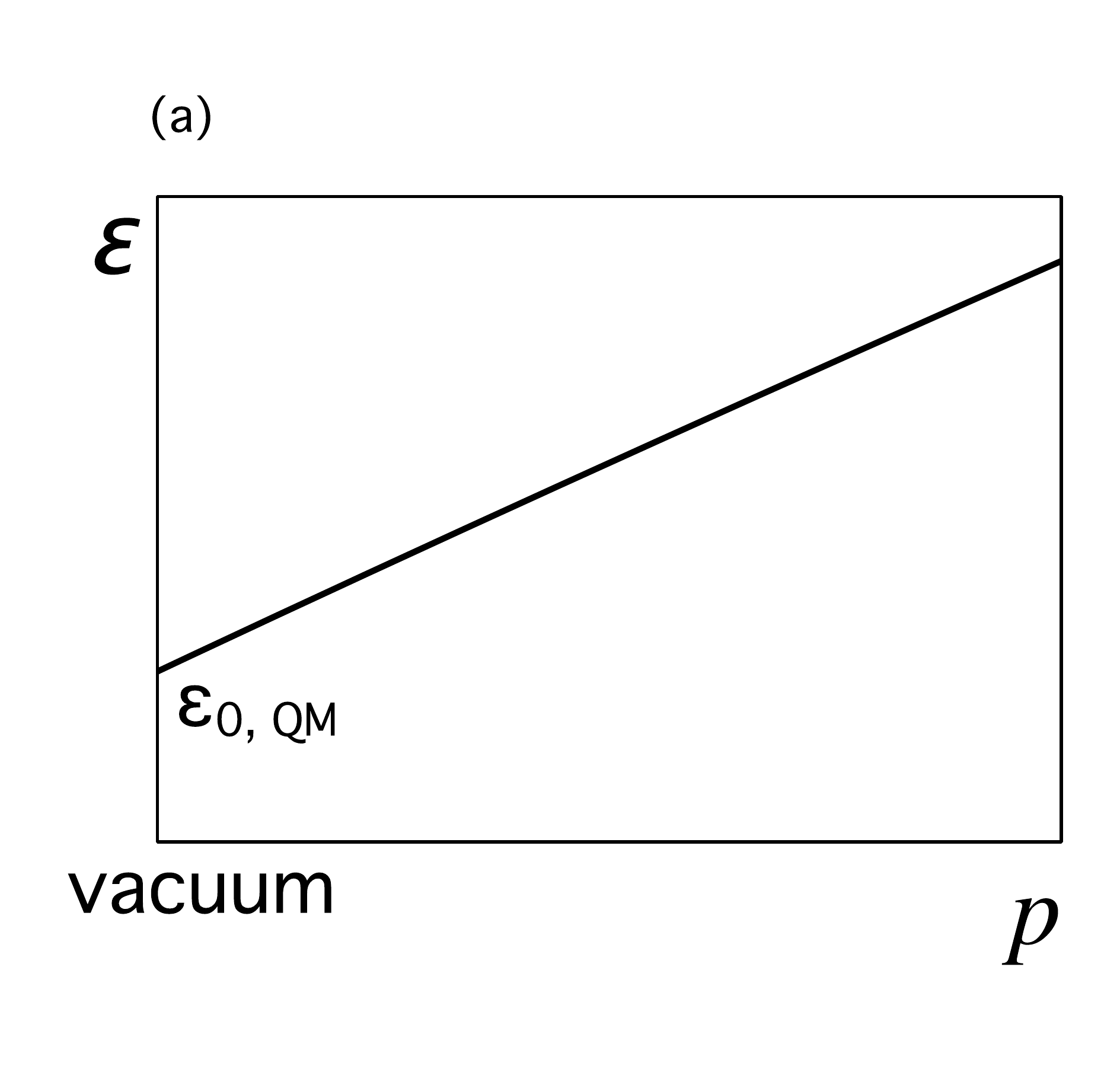}\\[-2ex]
}\parbox{0.33\hsize}{
\includegraphics[width=\hsize]{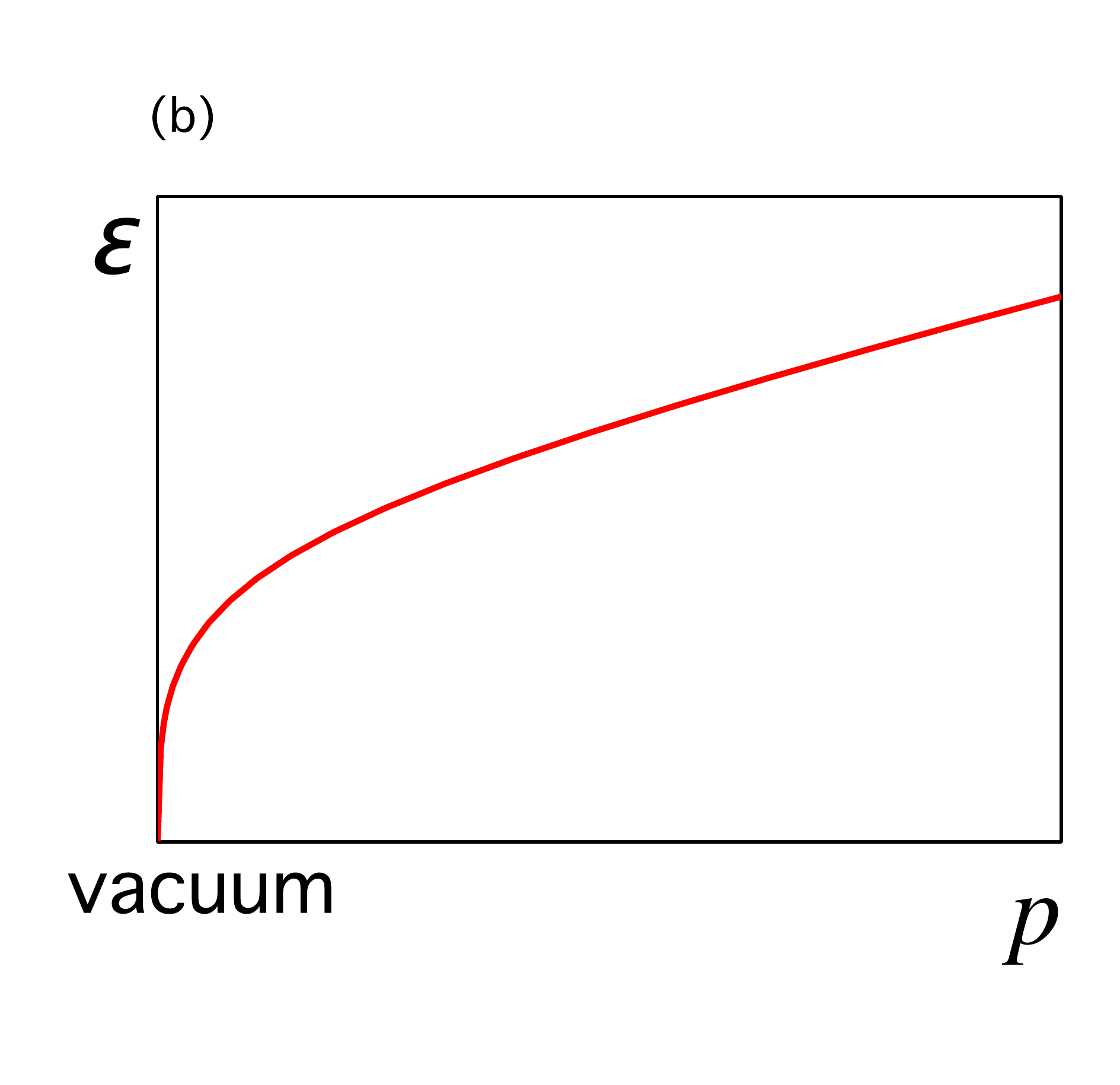}\\[-2ex]
}\parbox{0.33\hsize}{
\includegraphics[width=\hsize]{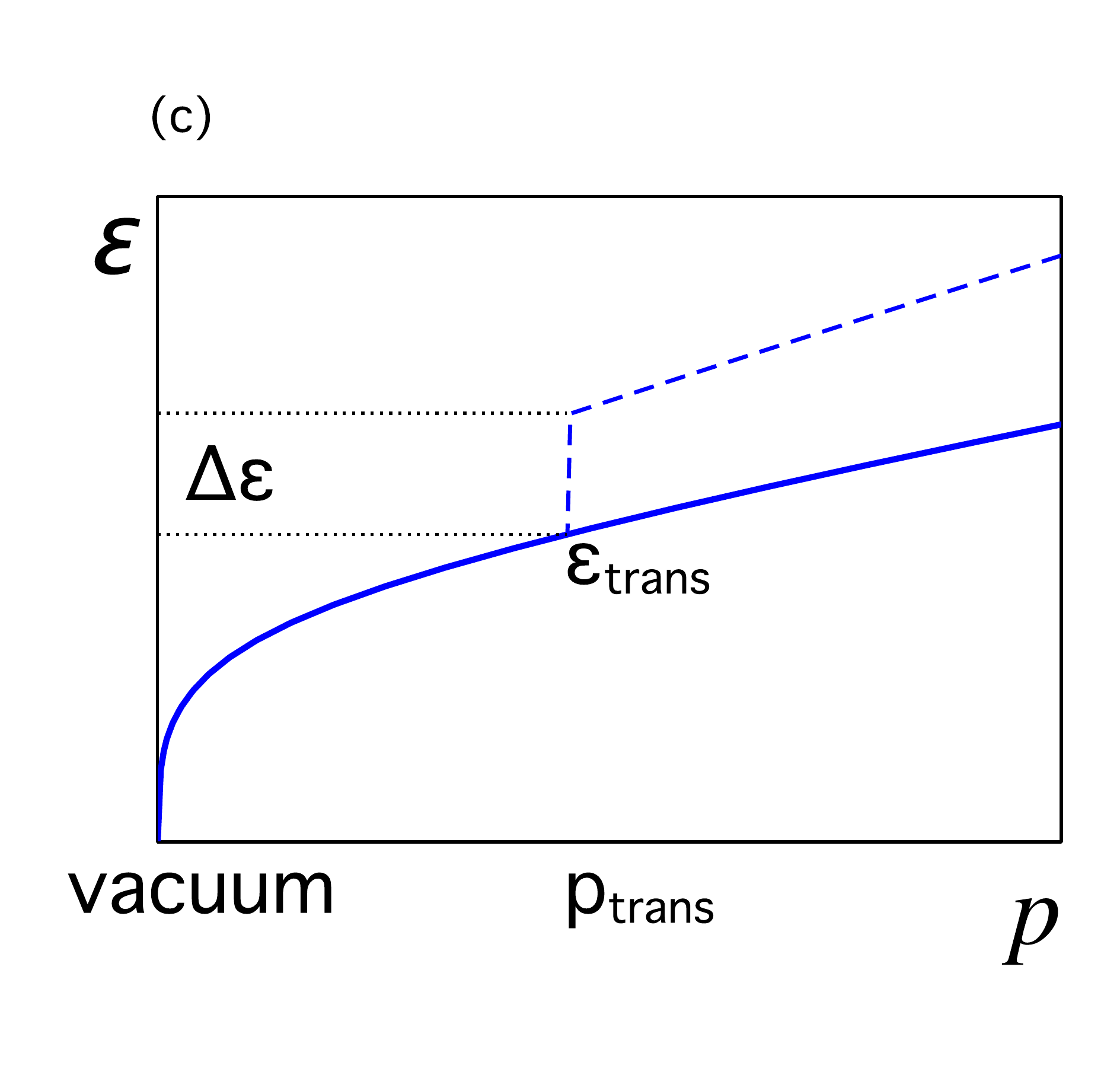}\\[-2ex]
}
\caption{Schematic EoSs that give rise to $M$-$R$ diagrams depicted in Fig.~\ref{fig:MR-illust}. For self-bound strange quark stars (panel (a)), the pressure becomes zero in equilibrium with vacuum at the surface with a finite density $\ep=\ep_{0, \rm QM}$. Panels (b) and (c) are consistent with experimental data of laboratory nuclei at saturation density $n_0$, but differ in their behaviors at higher densities. Parameters indicated in panel (c) characterize the onset density/pressure of a first-order phase transition ($\etrans$ and $\ptrans$) and its strength, $\De\ep$. The two hadronic EoSs used in the present paper (HLPS and DBHF) are consistent with modern chiral effective field-theoretical calculations~\cite{Drischler:2020hwi,Essick:2020flb} up to $\sim2\,n_0$. Note that the third-family stars in Fig.~\ref{fig:MR-illust} (c) are only possible for specific combinations of phase transition properties (see text).
}
\label{fig:EoS-illust}
\end{figure*}

In addition to the varying disposition of $\Rtyp$ with respect to $\Rtwo$, a survey of the trends of $M$-$R$ curves in Fig.~\ref{fig:MR-illust} reveals a few additional noteworthy points: 
\begin{enumerate}
\item For $M$-$R$ curves with $dR/dM > 0$ as in panel (b), a common radius exists for two different masses,\footnote{This feature is also reported for quarkyonic models, see e.g., Ref.~\cite{Zhao:2020dvu}.} and 
\item The third-family stars in panel (c) permit two different radii for the same mass.  
\end{enumerate}

Differentiating between these possibilities from observations clearly depends on the size of the difference in radii. However, these degeneracies raise an alert for analyses of data, whether from {\it{NICER}} or from GW measurements.

In the following, we elaborate on the case of hybrid stars with moderate to strong first-order phase transitions. For comparison, we also show typical results for self-bound and hadronic stars.

\section{The minimum radius of a $2\,\Msolar$ NS}
\label{sec:rmax}

For definiteness, we choose $2\,\Msolar$ as a proxy for a very massive star. The results shown below are not significantly affected by a slightly higher value. 

\subsection{Stiffness vs. compactness}

Self-bound strange quark stars without a crust behave uniquely on the $M$-$R$ diagram as the radius increases with mass until very close to the maximum-mass. 
Contrary to what is observed for purely-hadronic stars that higher the value of sound speed reached in matter (``stiffer'' EoS) the larger is the radius for a given mass, SQSs with decreasing sound velocity (``softer'' EoS) are less compact. The maximally compact configurations ~\cite{Lattimer:2012nd,Lattimer:2015nhk} of these objects are represented by a linear EoS 
\beq \varepsilon\left(p\right) =
4B^4+\frac{1}{\cQMsq} p \;
\label{eqn:eos_bare}
\eeq
that specifies the most stiff quark matter consistent with causality (squared sound speed $\cQMsq=1$) while supporting massive stars $\geq 2\,\Msolar$. 
Fixing $\Mmax=2\,\Msolar$ determines the remaining parameter, the bag constant $B$, to be about $186~\rm{MeV}$. Such an EoS leads to the smallest possible radius for a given mass, e.g., typical radius $\Rtyp= 8.09$~km, radius for the maximum-mass star $\Rmax=8.34$~km, minimum dimensionless tidal deformability $\La_{1.4}\simeq 59$, and also the lowest binary tidal deformability predicted $\tilde{\La}_{1.188}\in[64.97, 68.08]$ for a system with the same chirp mass and mass ratio range as measured in GW170817~\cite{Zhao:2018nyf,Han:2018mtj}.

The next competitive candidates for neutron stars with very high compactness, i.e., very small radii, are the so-called ``third-family'' stars \cite{Schertler:2000xq} built upon hybrid EoSs with a strong first-order phase transition at not-too-high densities in the supra-nuclear regime.  
A first-order transition with sufficient strength, characterized by the discontinuity in the energy density, $\De\ep/\etrans$, where $\etrans=\ep_{\rm HM}(\ptrans)$ stands for the energy density of the low-density hadronic matter at the transition, softens the EoS significantly leading to a drastic decrease ($\gtrsim2$~km) in radii for hybrid stars with higher central pressures~\cite{Alford:2013aca,Alford:2015gna}. 
If accompanied by a stiff enough high-density quark phase, the $\Mmax\geq2\,\Msolar$ constraint can be reasonably well satisfied (see e.g. Appendix~\ref{AppB}). On the other hand, if the transition strength is weak or the stiffness in quark matter is not significant, hybrid stars may have radii comparable to or even larger than those of normal hadronic ones. 
If, however, quarks appear through a smooth crossover instead of the commonly-assumed first-order transition, both the radii and maximum-masses are larger than those obtained from hadronic matter EoSs used due to an inherent boost in the pressure after quarks set in \cite{Kojo:2014rca,McLerran:2018hbz,Han:2019bub}.

To extensively study the prospects for future observational constraint on the minimum radius of high-mass neutron stars from x-ray missions, e.g., {\it{NICER}}, we concentrate on the first-order transition scenario with quark cores in neutron stars. Implications on the properties of the phase transition are deduced using the generic framework of the constant-sound-speed (CSS) parametrization~\cite{Alford:2013aca}:    
\beq \ep(p) = \left\{\!
\begin{array}{ll}
\ep_{\rm HM}(p), & p<\ptrans \\
\etrans+\De\ep+(p-\ptrans)/\cQMsq, & p>\ptrans
\end{array}
\right.\ 
\label{eqn:eos_hyb}
\eeq
The CSS parametrization is motivated by the observation that $\cQMsq$ is weakly dependent on density for a variety of widely-used quark matter EoSs. Examples include variations of the original MIT bag model~\cite{Baym:1976yu}, variations of the original Nambu-Jona--Lasinio (NJL) model~\cite{Nambu:1961tp}, perturbative quark  EoS~\cite{Kurkela:2009gj}, quartic polynomial parameterization~\cite{Alford:2004pf}, etc. 
Consequences of employing a realistic density-dependent behavior of $\cQMsq$ are easily gauged by examining results for neighboring values of $\cQMsq$. 
A verification that the static properties of NSs with density-dependent $\cQMsq$ lie between the cases of assuming constant $c_s^2$ equal to its minimum and maximal values can be found in~Refs.~\cite{Ranea-Sandoval:2015ldr,Ranea-Sandoval:2017ort}. 
For our purposes here, the CSS parametrization serves as a guide to capture the essential physics of quark matter.\footnote{Note that in quarkyonic matter, quarks are admixed with hadrons in a continuous crossover unlike in a pure quark phase. The speed of sound therefore varies with density in a very distinctive manner by exhibiting a prominent peak.} 
For normal hadronic matter, we select two representative models as baselines to gauge the dependence of outcomes on the low-density matter EoS. 
The HLPS EoS~\cite{Hebeler:2010jx} is soft at densities $\nb\lesssim 4\,n_{0}$, whereas DBHF \cite{GrossBoelting:1998jg} is a stiffer EoS, with higher pressure at a given energy density. 
The latter yields neutron stars that are larger, and can reach a higher maximum mass. We defer a detailed discussion on the choice of hadronic EoSs to Sec.~\ref{sec:had}, except to note here that both EoSs are reasonably consistent with modern chiral effective field-theoretical calculations up to $\sim2\,n_0$~\cite{Drischler:2020hwi,Essick:2020flb}. 
For NS crustal matter, we utilize the EoSs described in Refs.~\cite{Baym71tg,Negele73ns}. The crust-core transition density ($\sim 0.5\,n_{0}$) is determined by smoothly joining the low-density EoS of Ref.~\cite{Negele73ns} to the core EoSs used in this work. 
Use of the alternative low-density EoS of Ref.~\cite{haensel:1995eq} yields a similar transition density. The sub-nuclear density EoS, and differences caused by the above two low-density EoSs on massive stars are negligible, and in no way affect our conclusions. 

\begin{figure*}[]
\parbox{0.48\hsize}{
\includegraphics[width=\hsize]{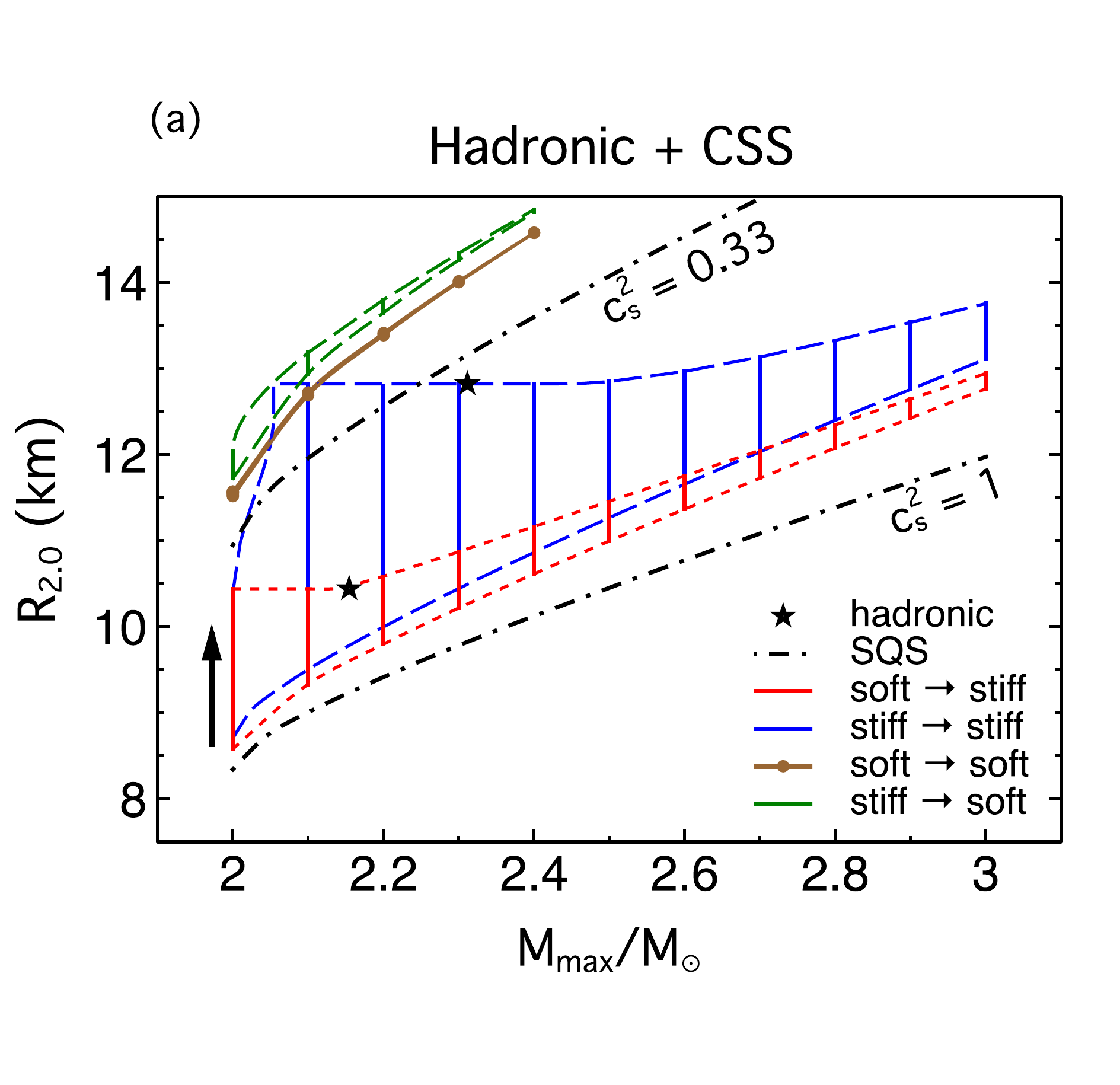}\\[-2ex]
}\parbox{0.48\hsize}{
\includegraphics[width=\hsize]{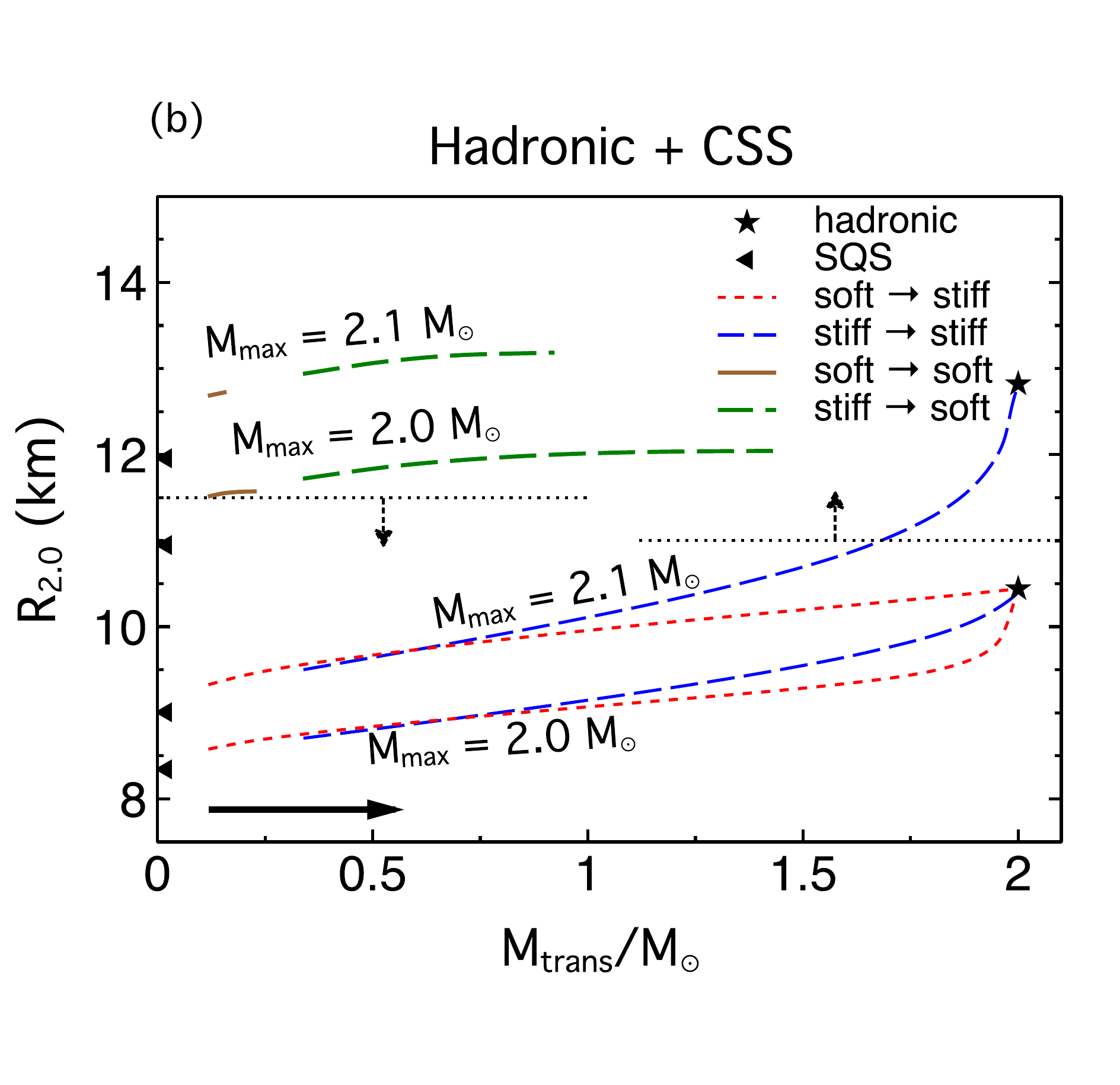}\\[-2ex]
}
\caption{Radius of a $2\,\Msolar$ star $\Rtwo$ vs. maximum mass $\Mmax$ (left panel) and $R_{2.0}$ vs. mass of the heaviest hadronic star $\Mtrans$ for a fixed $\Mmax=2.0 \,\Msolar, 2.1 \, \Msolar$ (right panel). Hybrid stars were constructed through a first-order transition with a sharp interface using the generic CSS parametrization. The two representative hadronic matter EoSs chosen are HLPS (soft) and DBHF (stiff).
The squared speed of sound $\cQMsq$ in quark matter varies between 0.33 and 1. For comparison, results for purely-hadronic EoSs (asterisks) and SQSs are also shown. 
In the left panel, solid vertical lines refer to a fixed value of $\Mmax$ along which the corresponding transition density, $\ntrans$, above which quark matter emerges, or the onset mass, $\Mtrans$, of phase transition in parallel (right panel $x$-axis) are increasing upwards. 
The minimum value of $\Rtwo$ on each vertical line refers to the lowest possible $\ntrans=n_{0}$, whereas the maximum value of $\Rtwo$ represents the highest possible $\ntrans$, which is determined by either requiring a finite strength of the phase transition $\De\varepsilon>0$ or an artificial cutoff on $\Mtrans<2\,\Msolar$. 
In the right panel, two hypothetical observational constraints, $\Rtwo\leq11.5$~km and $\Rtwo\geq11$~km, are also displayed (dotted horizontal lines with arrows).
}
\label{fig:R20_base}
\end{figure*}

In Fig.~\ref{fig:R20_base}, we plot the radius $\Rtwo$ of a $2\,\Msolar$ star from solutions to the TOV equation of hydrostatic equilibrium~\cite{Tolman:1939jz,Oppenheimer:1939ne} for different hybrid EoSs together with $\Mmax$ (left panel), and $\Mtrans$, the mass of the heaviest hadronic star (right panel) from the same EoS. 
Results for SQSs and purely-hadronic stars are also shown for comparison. To explore the largest parameter space available, we have chosen quark matter being either very soft or maximally stiff, with squared sound velocity $\cQMsq=0.33$ and $1$, respectively. The former value is characteristic of very weakly-interacting relativistic quarks, whereas the latter is set by the limit of causality.

\begin{figure*}[]
\parbox{0.48\hsize}{
\includegraphics[width=\hsize]{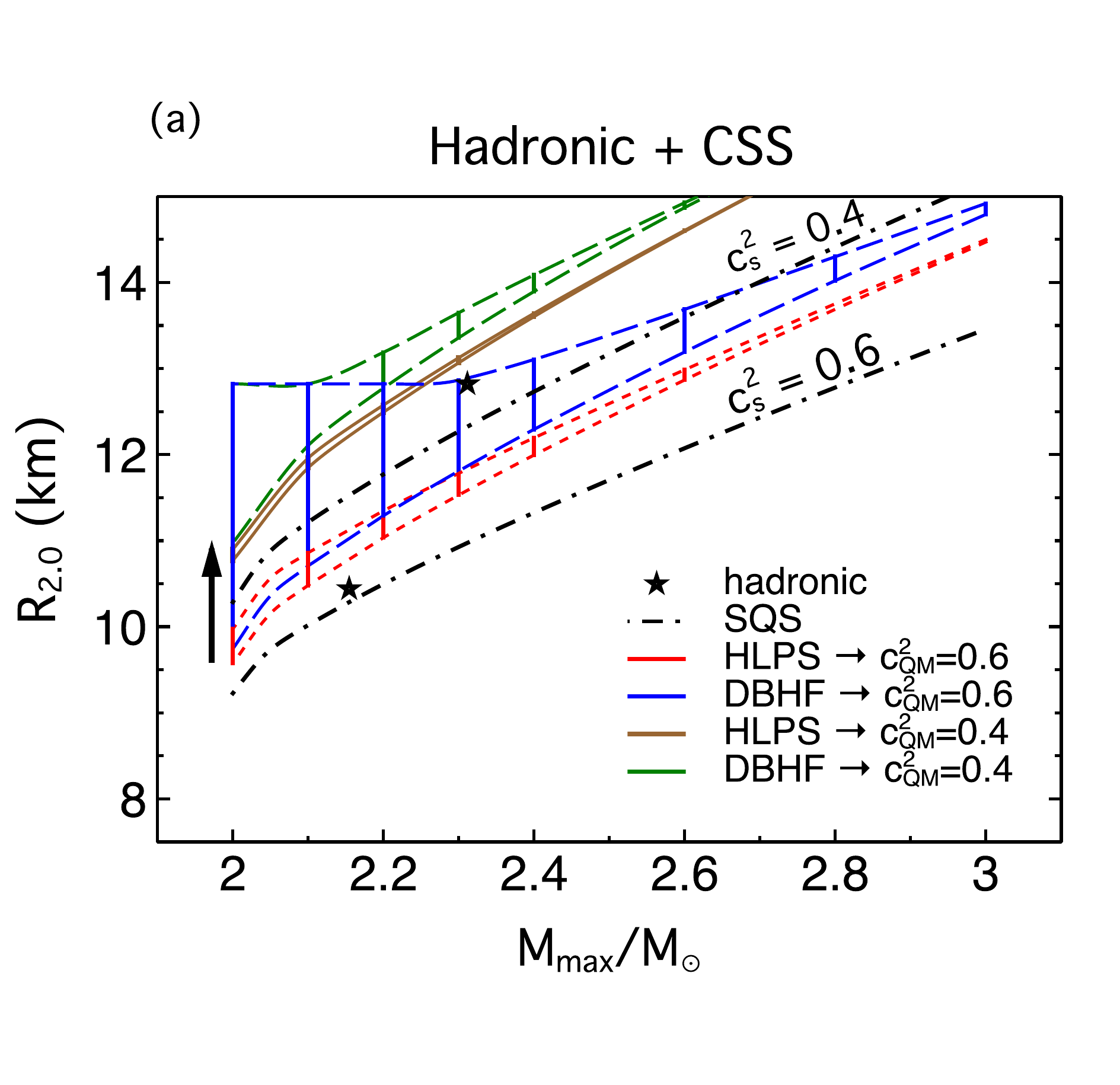}\\[-2ex]
}\parbox{0.48\hsize}{
\includegraphics[width=\hsize]{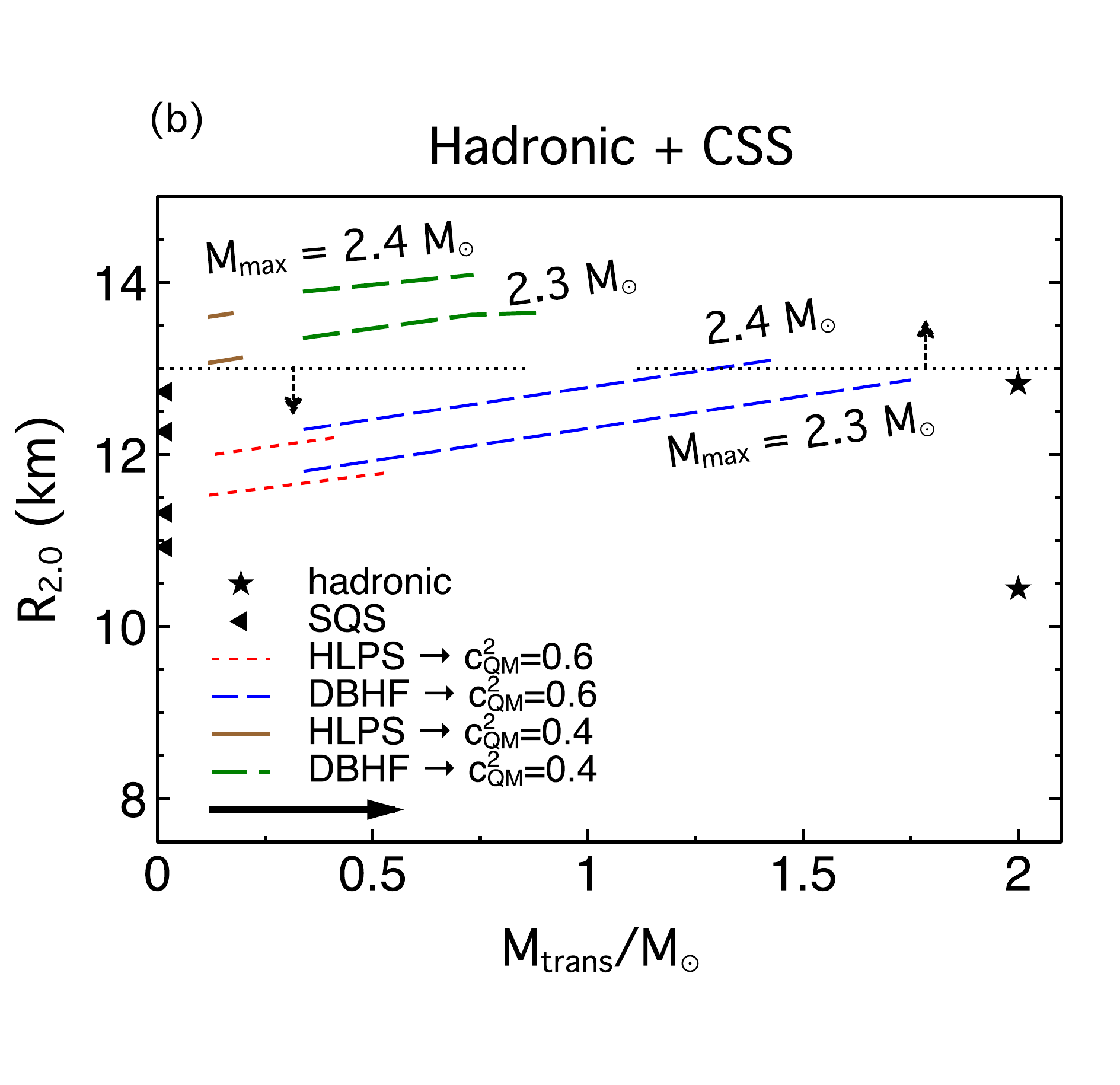}\\[-2ex]
}
\caption{Same as Fig.~\ref{fig:R20_base}, but with different sound speed values in quark matter for hybrid stars ($\cQMsq=0.4, 0.6$), and accordingly for self-bound stars ($c_s^2=0.4,0.6$). In the right panel, maximum masses are fixed at the higher values of $2.3$ and $2.4\,\Msolar$, and confronted with hypothetical radius bounds from observation, $\Rtwo\leq13$~km or $\Rtwo\geq13$~km (dotted horizontal lines with arrows).
}
\label{fig:R20_more}
\end{figure*}

\begin{figure*}[]
\parbox{0.48\hsize}{
\includegraphics[width=\hsize]{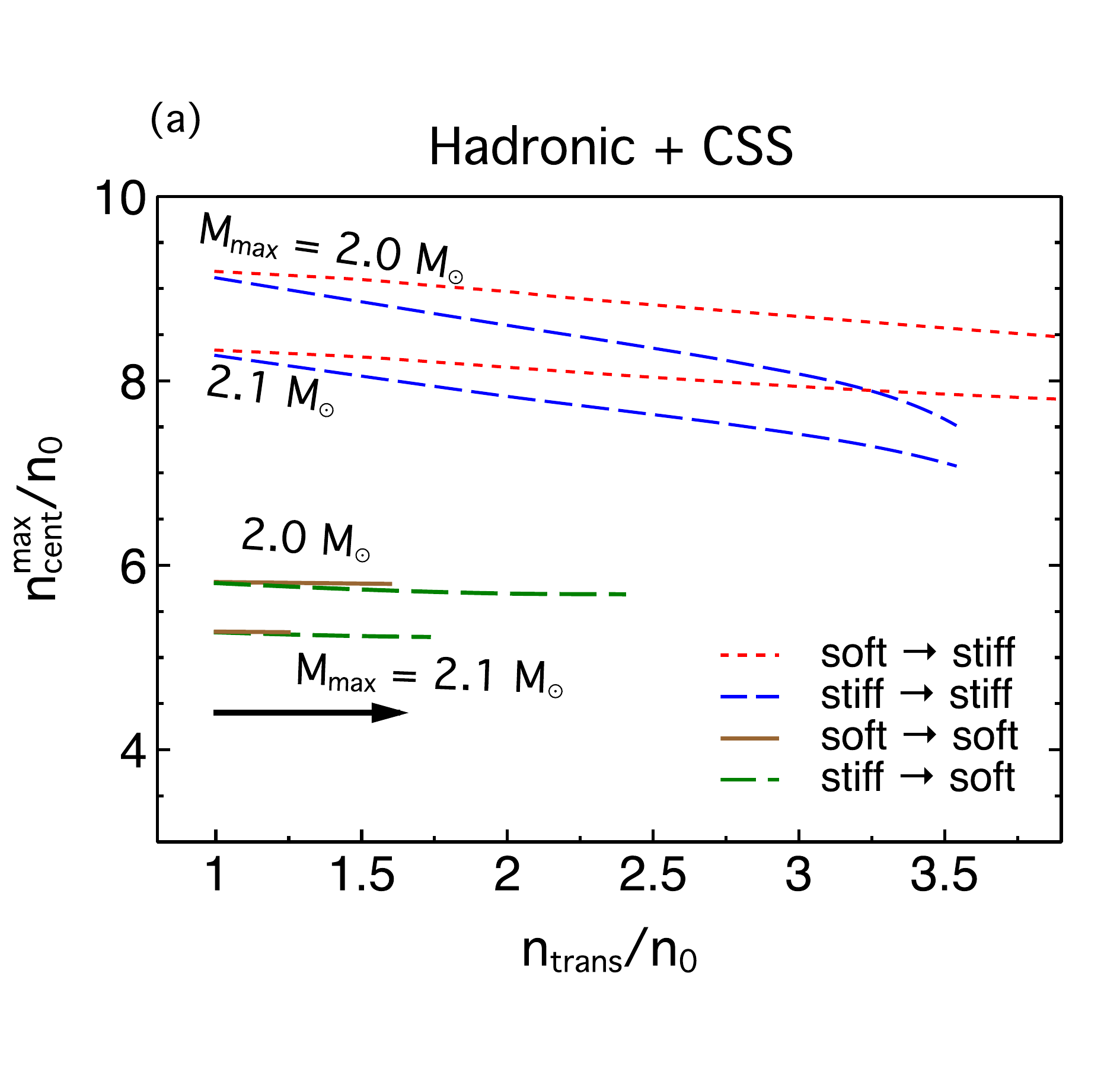}\\[-2ex]
}\parbox{0.48\hsize}{
\includegraphics[width=\hsize]{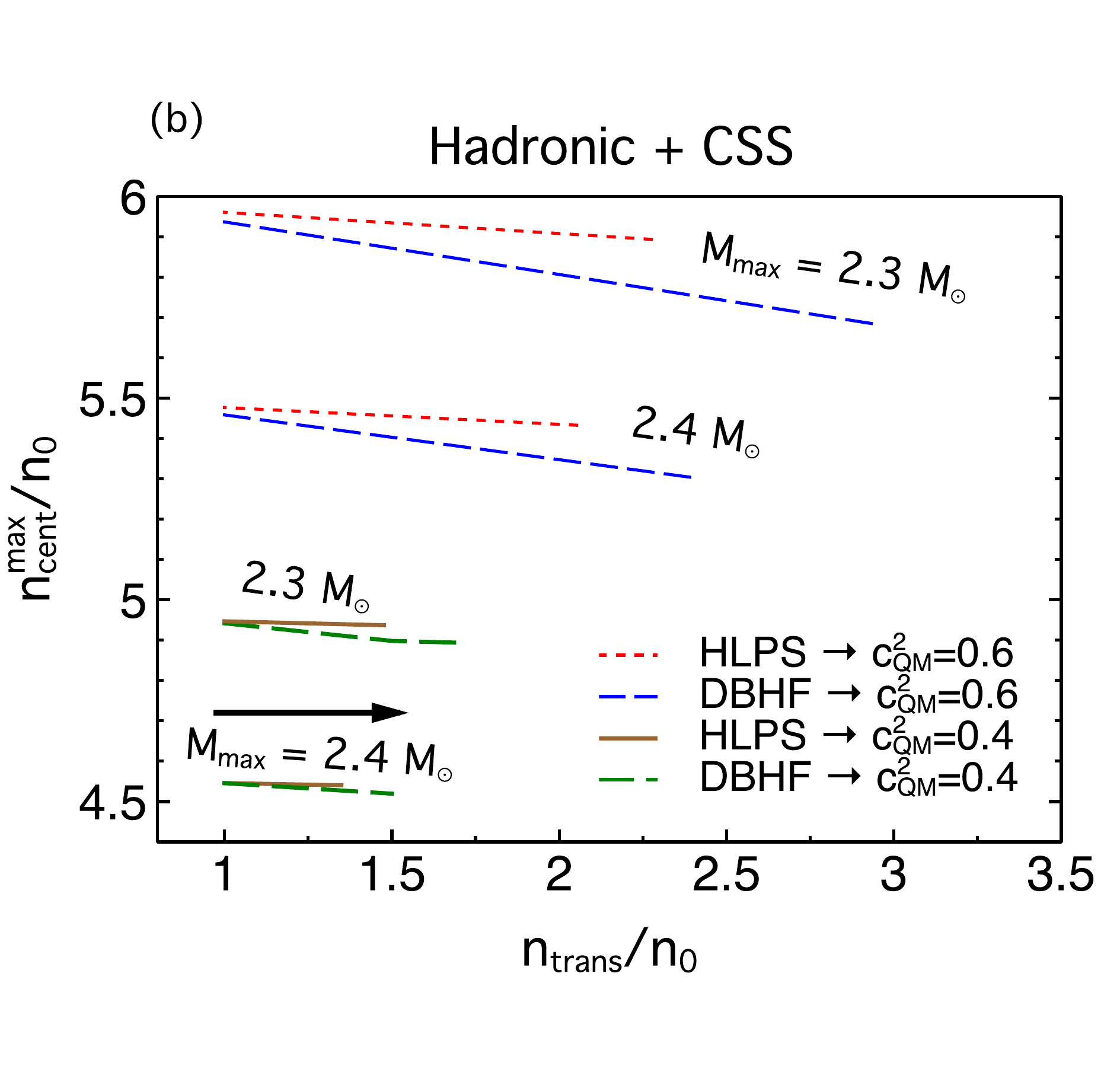}\\[-2ex]
}
\caption{Central density of the maximum-mass star $\nmax$ vs. 
the transition density $\ntrans$. In the left panel, the parameter space of the hybrid EoSs is the same as in Fig.~\ref{fig:R20_base} (b). The red dotted lines extend to high densities of $\approx 5\,n_{0}$ that are outside of the plot. In the right panel, the parameter space of the hybrid EoSs are as in Fig.~\ref{fig:R20_more} (b). Results for $\ntrans /n_0 \lesssim 2$ are to be viewed as extreme cases as they would be in conflict with nuclear matter studies and heavy-ion phenomenology (see text).
}
\label{fig:nmax_ntrans}
\end{figure*}

As mentioned before, the ``maximally compact'' EoS for self-bound stars with $\cQMsq=1$ gives rise to the smallest radius for a given mass, and for a $2\,\Msolar$ star this value is $8.34$~km if the maximum mass is also set equal to $2\,\Msolar$. 
With a higher maximum mass, obtained by setting a smaller bag constant $B$ in \Eqn{eqn:eos_bare}, $\Rtwo$ becomes larger. Similarly, for very soft quark matter ($\cQMsq=0.33$), the smallest $\Rtwo$ of SQSs is $10.95$~km with $\Mmax=2\,\Msolar$. 
For hybrid stars, the results are more complicated because the low-density input from hadronic matter may also influence the radius, particularly when the phase transition only takes place at high enough densities. 
In the left panel, the four colored bands with vertical lines inside indicating different values for the fixed $\Mmax$ correspond to the following four combinations of hadronic and quark matter in phase transitions: 
\begin{enumerate}

\item Soft (HLPS) $\to$ stiff ($\cQMsq=1$); red on plot. This set of parameters reaches the most compact configurations when the transition occurs at $\ntrans=n_{0}$, represented by the lower boundary that starts with $\Rtwo=8.58$~km when $\Mmax=2\,\Msolar$ and increases afterwards with higher maximum mass. 

If the transition occurs for $\ntrans> n_{0}$, indicated by the upward arrow along the vertical lines, then $\Rtwo > 8.58$~km until at some density when there is no valid phase transition (i.e. $\De\ep=0$) or it reaches the central density of a hadronic $2\,\Msolar$ star; see contours plots on the right panel.

\item Stiff (DBHF) $\to$ stiff ($\cQMsq=1$); blue on plot. This set covers the widest range of radius, mainly because of the largest available parameter space for $\ntrans$ (with valid strength $\De\ep$), and also because the hadronic baseline being stiff points to a larger radius before the transition happens. In most cases, $\Rtwo$ of hybrid stars are smaller than their hadronic counterpart $\Rtwo^{\rm DBHF}=12.82$~km, but it is still possible that some configurations have even larger values if $\Mmax\gtrsim2.5\,\Msolar$. 
The heavy NS candidate for the  secondary component with $\sim 2.6\,\Msolar$ of GW190814~\cite{Abbott:2020khf} consistent with GW170817 can be readily obtained in this scenario if the transition takes place at low enough densities ($\lesssim 2.8\, n_{0}$assuming maximally stiff quark matter). Reducing $\cQMsq$ pushes available $\ntrans$ into lower values, see Fig.~\ref{fig:R20_more} (a).

\item Soft (HLPS) $\to$ soft ($\cQMsq=0.33$); brown on plot. The most severely restricted parameter space is usable for this combination because the allowed values of $\ntrans$ barely vary, only acceptable at $\ntrans\lesssim1.5\,n_{0}$; see contours plots on Fig.~\ref{fig:nmax_ntrans} (a).

\item Stiff (DBHF) $\to$ soft ($\cQMsq=0.33$). The overall $\Rtwo$ values indicated are larger compared to other sets, and the stiffer hadronic part help maintain valid transitions to slightly higher densities above saturation $1.75-2.4\, n_{0}$ if $\Mmax\leq2.1\,\,\Msolar$. 
An interesting feature here is that the restrictions imposed by the allowed ranges of $\ntrans$ constrain the sound speed in nuclear matter to be typically small, with $\cNMsq<1/3$ (which is also the case for the soft HLPS EoS above). 
Combined with very soft quark matter ($\cQMsq\approx0.33$) assumed at higher densities, the overall hybrid EoS with a first-order phase transition obeys the bound $c_{s}^2 \leq 1/3$ consistent with conformal theories or perturbative QCD \cite{Borsanyi:2012cr,Kurkela:2014vha,Bedaque:2014sqa}. 
In contrast, it is nearly impossible for a normal EoS without discontinuities to accommodate massive neutron stars of $\gtrsim 2\,\Msolar$ with the squared sound speed staying $c_s^{2}\leq1/3$~\cite{Moustakidis:2016sab,Tews:2018kmu}.
\end{enumerate}

Regardless of the stiffness in either hadronic or quark matter, the minimum $\Rtwo$ is always associated with the lowest possible $\ntrans=n_{0}$ for a fixed maximum mass (but still above the bound set by ``maximally compact'' SQSs). 
As $\ntrans$ increases (so does the mass of the heaviest hadronic star $\Mtrans$), $\Rtwo$ also increases. The maximal $\Rtwo$ is then limited by when the transition strength drops to zero ($\De\ep=0$), or the heaviest hadronic star is already at $2\,\Msolar$ which is imposed as an artificial cutoff in the present study. 
To better visualize the trend, we plot selected maximum-mass contours in Fig.~\ref{fig:R20_base} (b); example $M$-$R$ curves for $\Mmax=2.0\,\Msolar$ are also shown in Appendix~\ref{AppA}. 
Note that in the stiff $\to$ stiff case with $\Mmax=2.0\,\Msolar$, $\Rtwo$ ends at the same value as for the soft hadronic EoS. That the third-family branch deviating from the stiffer DBHF EoS happens to intersect with the hadronic branch of soft HLPS EoS at $\Mmax=2.0 \,\Msolar$ is a coincidence; see also the inset plot of Fig.~\ref{fig:MR_Mmax20} (a).
 
Figure~\ref{fig:R20_base} (b) also demonstrates how hypothetical radius measurements, with an upper or lower bound on a $2\,\Msolar$ star (horizontal dotted lines) would place constraints on the hybrid EoSs. 
If $\Rtwo<11.5$~km were to be established, then the lower bound $\Mmax>2\,\Msolar$ implies that weakly interacting quarks with $\cQMsq\lesssim 1/3$ are convincingly ruled out except for SQSs. Alternatively, with a lower bound $\Rtwo>11$~km and an upper bound $\Mmax<2.1\,\Msolar$ (hypothetically), the scenario that a sharp transition into maximally stiff quark matter takes place in $\lesssim 1.67\,\Msolar$ neutron stars is strongly disfavored.

It is not surprising that an upper bound on the radius of high-mass stars would be more restrictive, not only because a small $\Rtwo$ typically requires a low $\ntrans$ (which diminishes the complicating effect from the hadronic baseline assumed), but also because the current lower bound on $\Mmax$ (primarily from the heavy pulsars detected and measured to good precision) is more reliable than the upper bound on $\Mmax$ (inferred from various analyses that involve simulations and modeling with larger uncertainties).

Figure~\ref{fig:R20_more} shows results for $\Rtwo$ and $\Mmax$ contours with the same HLPS (soft) and DBHF (stiff) EoSs for hadronic matter, but with quark EoSs with moderate stiffness, $\cQMsq=0.4$ and $\cQMsq=0.6$, respectively. The SQS EoSs are also modified such that they contain quark matter of the same stiffness. In addition, a bolder assumption on the maximum mass $\Mmax=2.3$ or $2.4\,\Msolar$ is specified on the right panel, Fig.~\ref{fig:R20_more} (b). 

Following the same argument as for the previous figure, future radius measurements of $2\,\Msolar$ stars can be used to infer information on properties of the phase transition. If $\Rtwo<13$~km were to be preferred, then combining the (hypothetical) lower bound on the maximum mass $\Mmax>2.3\,\Msolar$ soft quark matter with $\cQMsq\lesssim 0.4$ can be almost excluded except for SQSs. Alternatively, with a lower bound $\Rtwo>13$~km and an upper bound $\Mmax<2.4\,\Msolar$, it is unlikely that hadronic matter undergoes a sharp transition into quark matter with $\cQMsq\gtrsim 0.6$ below the central density of a $1.44\,\Msolar$ star. It is worth noting that higher values of $\Mmax$ necessitates lower $\ntrans$. Sometimes the transition mass $\Mtrans\lesssim 1.1\,\Msolar$ happens to be even smaller than the lowest NS mass measured \cite{Martinez:2015mya} or suggested by astrophysical evolutionary scenario~\cite{Lattimer:2012nd,Suwa:2018uni}, indicating all NSs observed are actually hybrid stars.

\begin{figure*}[]
\parbox{0.48\hsize}{
\includegraphics[width=\hsize]{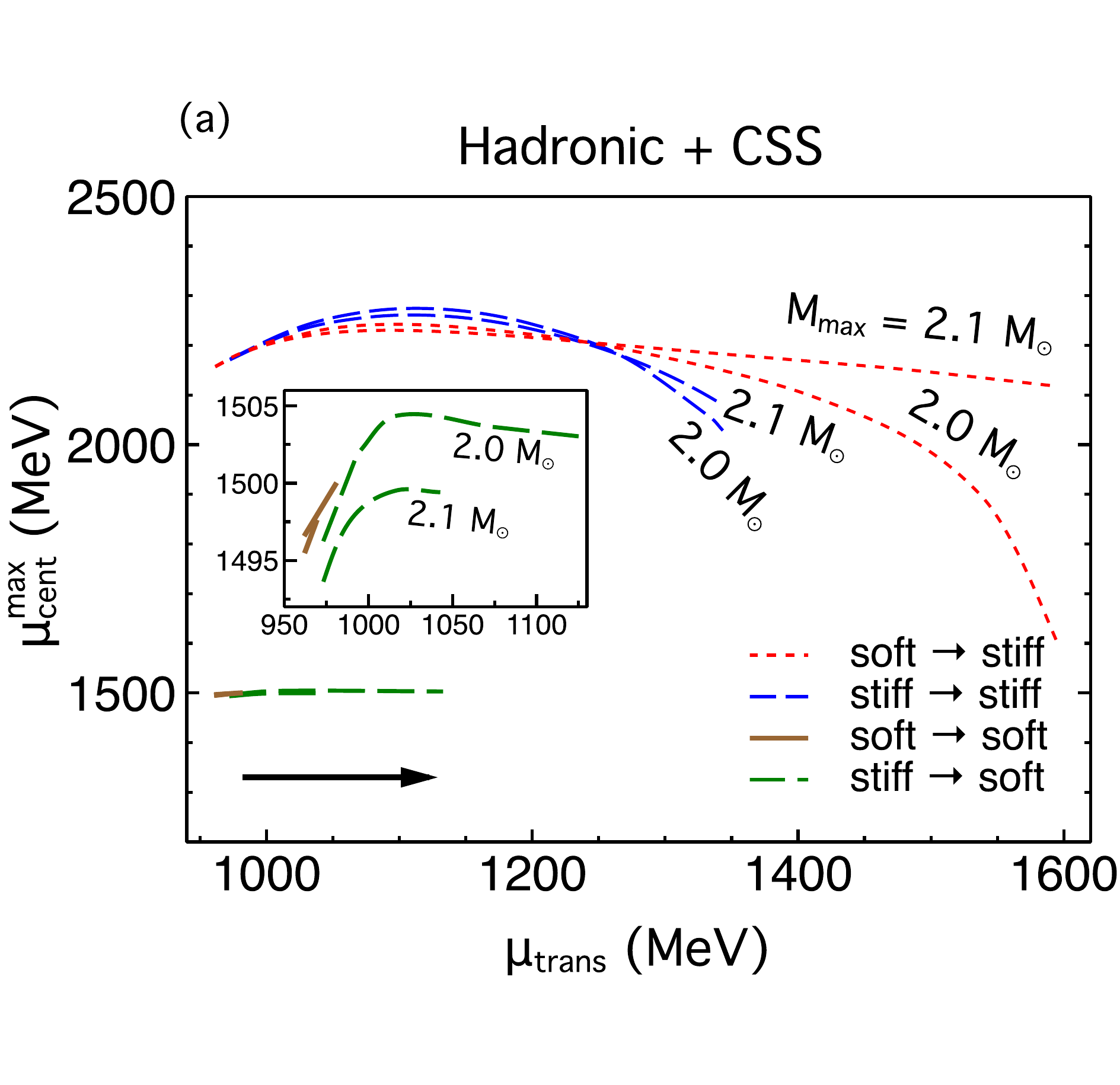}\\[-2ex]
}\parbox{0.48\hsize}{
\includegraphics[width=\hsize]{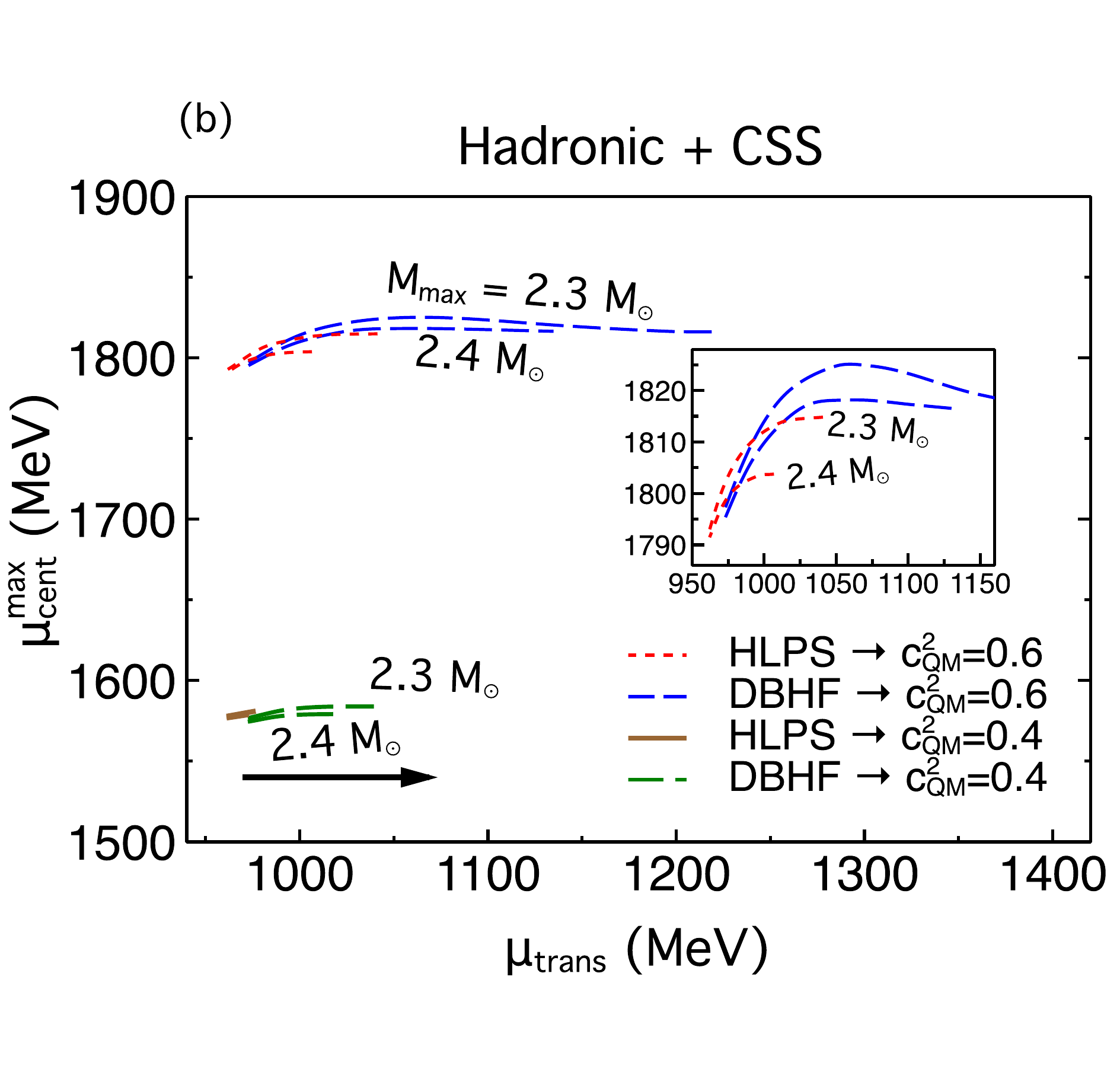}\\[-2ex]
}
\caption{Highest chemical potential reached in the maximum-mass star, $\mumax$, as a function of the critical chemical potential at transition, $\mutrans$. EoS parameters are the same as those in Fig.~\ref{fig:nmax_ntrans}. 
}
\label{fig:mumax_mutrans}
\end{figure*}

\begin{table*}[htb]
\begin{center}
\begin{tabular}{c c@{\quad} c@{\quad} c@{\quad} c@{\quad} c@{\quad} c@{\quad} c@{\quad} c@{\quad} c@{\quad} c@{\quad} c@{\quad}  c@{\quad}  c@{\quad}  c}
\hline \hline
property & $p_{\rm sat}$ &  $L$ & $\Mmax$  & $\Rmax$  &  $\csqmax$  & $\nmax$ & $\Rtwo$ &  $\csqtwo$  & $\ncent^{2.0}$ & $\Rtyp$ & $\csqtyp$  & $\ncent^{1.4}$ & 
$c_{s,1.1}^2$ &  $\ncent^{1.1}$ \\
 units & ($\rm{MeV/fm^{3}}$)  &  (MeV)     & ($\Msolar$) & (km) &   &  ($n_0$) &   (km)  &    &  ($n_0$)   & (km) &     & ($n_0$)    &    &   ($n_0$)  \\
\hline
HLPS &  1.589  &  33.0    & 2.15 & 9.58  & 1.512  & 7.6225 &  10.44  & 1.012
 &  5.503 & 10.88 & 0.522 & 3.771 & 0.383 & 3.239  \\
\hline
DBHF & 3.982  &  69.4 & 2.31 & 11.26   & 0.999   & 6.0056  &  12.82 &  0.571 
& 3.536
 & 13.41 & 0.314 & 2.363 & 0.236 & 1.958 \\
\hline
\end{tabular}
\end{center}
\caption{Calculated properties for the hadronic EoS baselines used in this work. HLPS is softer, DBHF is stiffer. 
The symmetry energy slope parameter $L= 3n_0(\p S_0/\p n)|_{n_0}$ (where the symmetry energy $S_0$ refers to the difference of the energies of symmetric nuclear matter and pure neutron matter), and the pressure of nuclear matter at saturation density $p_{\rm sat}$ are constrained from low-energy nuclear theory and laboratory experiments. 
The maximum-mass star with $M=\Mmax$ is also the most compact, with its radius being the smallest $R=\Rmax$. Values of the central density $\ncent$, sound speed squared $c_s^2$ at the central density, and the radius are displayed for various masses $\Mmax$, $2.0 \,\Msolar$, $1.4\,\Msolar$ and $1.1\,\Msolar$ (except $R_{1.1}$ which is close to $\Rtyp$). 
Note that the sound speed in HLPS model rises above 1 and becomes unphysical at $\approx 5.5\,n_0$, $p \approx 405\, \rm{MeV/fm^{3}}$, which also corresponds to the central density of $M\approx 2\, \Msolar$ star.
}
\label{tab:had_EoS}
\end{table*}  

In Figs.~\ref{fig:nmax_ntrans} and~\ref{fig:mumax_mutrans}, we plot out the ranges of baryon density and chemical potentials\footnote{The chemical potential, $\mutrans$, is determined by the standard rules of chemical potential and pressure equalities for a Maxwell construction. 
Explicitly, $\mutrans = \mu_n = \sum_{i=n,p,e^-, \mu^- } Y_i \mu_i$ (where $Y_i$ are the particle concentrations), which assures that compositional details are preserved below $\ntrans$ and for the entire density range in which a mixed phase is present. 
However, details about the composition are lost for $\nb > \ntrans$ in the CSS- and other similar parametrizations of the pure quark phase.
} 
covered in the hybrid EoSs in Fig.~\ref{fig:R20_base} (b) and Fig.~\ref{fig:R20_more} (b). 
Results of $\nmax$ for values  of $\ntrans/n_0 \lesssim 2$ in these figures should be considered as extreme cases as such low transition densities would be in conflict with nuclear matter studies~\cite{Drischler:2020hwi,Essick:2020flb}, and the analysis of flow observables in heavy-ion collisions in the energy range up to 4 GeV per particle beam energy with nucleonic degrees of freedom only~\cite{Danielewicz:2002pu}.  Nonetheless, they provide conservative constraints with values not too different from those for $\ntrans/n_0 \gtrsim 2$.

Increases in $\ntrans$ or the critical chemical potential $\mutrans$ for the phase transition (along the $x$-axis with arrow rightwards) is reflected as rising $\Rtwo$ (along the $y$-axis with arrow upwards) in Fig.~\ref{fig:R20_base} (a) and Fig.~\ref{fig:R20_more} (a). 
The central density and chemical potential attainable are mostly sensitive to the stiffness in quark matter, with the highest $\nmax\approx9\,n_{0}$ and $\mumax\approx 2300$~MeV for $\cQMsq=1$ and $\Mmax=2\,\Msolar$, although there is a decreasing behavior of $\nmax$ with increasing $\ntrans$ which is more notable when the low-density hadronic matter is stiffer. 
On the other hand, the maximal $\ntrans$ varies for several reasons. If $\ntrans$ approaches high enough density close to the center of a $2\,\Msolar$ hadronic star (which is possible if the maximum mass is not far above $2\,\Msolar$ and quark matter is sufficiently stiff), $\ntrans$ is also higher with a softer hadronic EoS as a softer hadronic EoS reaches larger central densities for a given mass. 
If $\ntrans$ has to be truncated at lower densities (forced by the requirement to ensure a valid phase transition $\De\ep>0$), then a stiffer hadronic EoS normally leads to a higher maximal $\ntrans$. The maximum mass also exerts a more profound influence on $\nmax$ compared to $\mumax$, and on the maximal $\ntrans$ than the maximal $\mutrans$. 

It has been pointed out by Ref.~\cite{Radice:2016rys} that an unusual (but smooth) softening due to hyperons in the EoS may change the qualitative dynamics (such as in the maximum density and binding energy) of the remnant evolution in binary mergers. 
It would be instructive to investigate consequences for EoSs with substantial softening induced by sharp phase transitions that can access maximally high density, similar to what have been illustrated in the current work.

The critical chemical potential of a first-order transition $\mutrans$ on the conjectured quantum chromodynamics (QCD) diagram is unknown and cannot be computed from first-principle perturbative theories. The potential constraints on $\mutrans$ and $\mumax$ from inferred NS maximum mass shown in Fig.~\ref{fig:mumax_mutrans} therefore might be useful to connect to other regions probed on the QCD phase diagram, e.g., through a switching function relating to heavy-ion collisions at high temperature \cite{Plumberg:2018fxo} or extrapolation from perturbative QCD at ultra-high density \cite{Annala:2019eax}.

Theoretical interpretations of low-to-intermediate energy ($0.5-4$ GeV beam energy per particle) nearly isospin symmetric heavy-ion collisions offer another source for probing the dense matter EoS for densities up to $(3-4)\,n_0$~\cite{Danielewicz:2002pu}.  Analyses of such data in the above energy range using  Boltzmann-type kinetic approaches, in which effects of both mean fields (of relevance to the equilibrium EoS) and collisions are considered, have largely been with nucleonic degrees of freedom. It would be interesting and desirable to extend such an approach to include
quark degrees of freedom and their subsequent hadronization as in RHIC and CERN experiments at higher energies. The challenge here lies in disentangling the role of mean field effects from those involving collisions, and extrapolation to the neutron-rich conditions in neutron stars.

\subsection{Sensitivity to the hadronic EoS}
\label{sec:had}

Throughout our calculations, we have excluded transition to quark matter occurring below nuclear saturation density ($\ntrans < n_0$) as otherwise bulk nuclear matter would be metastable. 
The central density of the heaviest hadronic star, $\nmax$, places an upper limit on $\ntrans$, although in this study (for simplicity) we have employed an artificial termination at $\Mtrans\leq 2\,\Msolar$ which translates into different values of $\ntrans$ for different hadronic models. 
For a given mass, $\ncent$ of the NS is higher if the EoS is softer, and the largest value of $\ntrans$ limited by $\Mtrans\leq2\,\Msolar$ is $\approx 5.5\,n_0$ for HLPS and $\approx 3.54 \,n_0$ for DBHF. 
Table~\ref{tab:had_EoS} briefly summarizes the essential properties of these two EoSs and the resulting hadronic stars. 

\begin{figure}[]
\includegraphics[width=\hsize]{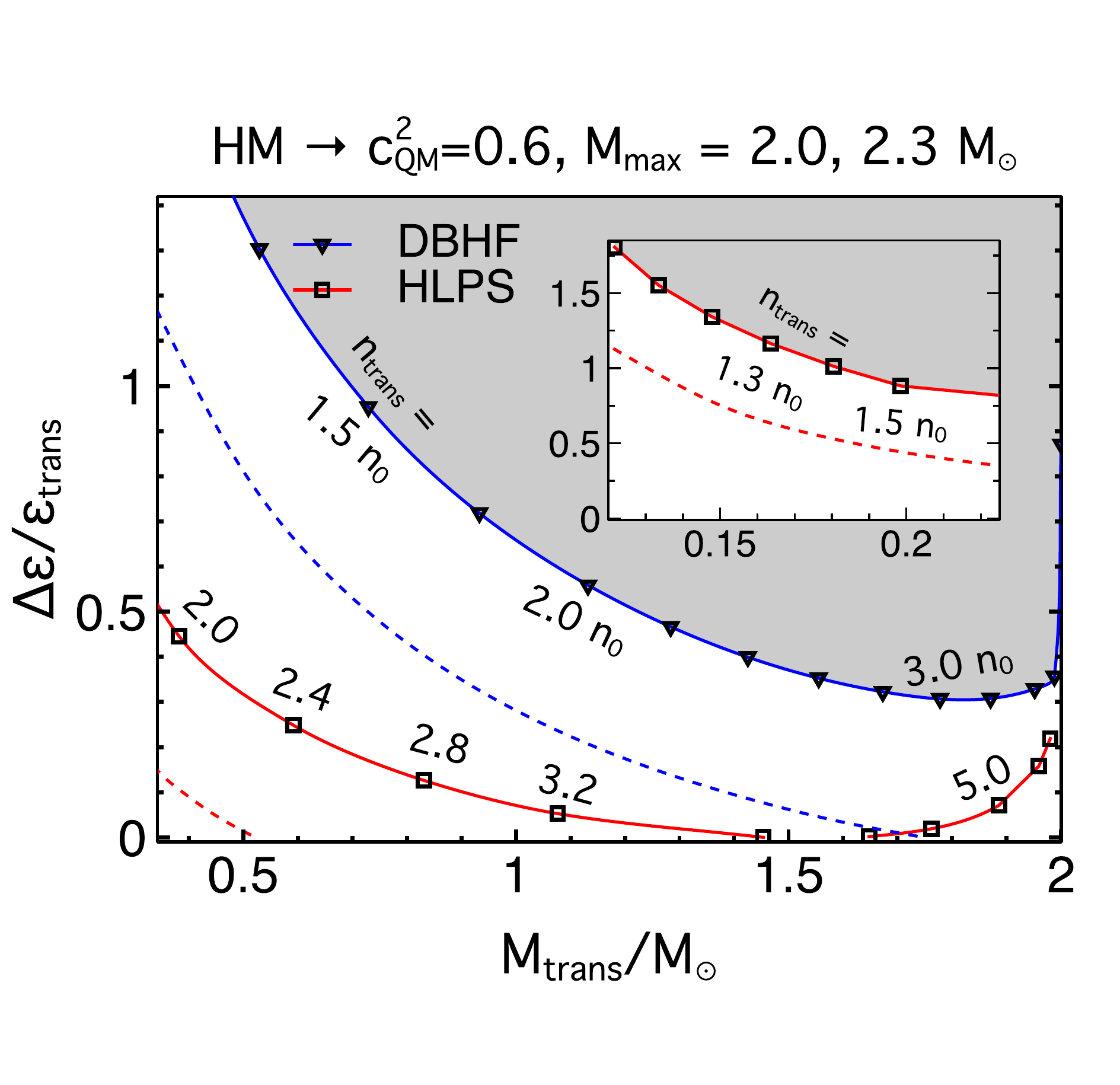}\\[-4ex]
\caption{
Relation between the phase transition strength $\De\ep/\etrans$ and its onset mass $\Mtrans$ for hybrid stars with $\Mmax=2.0\,\Msolar$ (solid) and $\Mmax=2.3\,\Msolar$ (dotted), obtained for the HLPS and DBHF hadronic EoSs with relatively stiff ($\cQMsq=0.6$) quark matter. 
Typical values of the transition density $\ntrans$ are specified along each contour. The gray-shaded regions on the upper-right corner mark the area excluded by heavy pulsar measurements $\Mmax\geq 2.0\,\Msolar$ for DBHF (inset for HLPS with very small values of $\Mtrans$). The detection of even heavier pulsars will further reduce the allowed parameter space.
}
\label{fig:Mt_Dt_Mmax}
\end{figure}

Figure~\ref{fig:Mt_Dt_Mmax} illustrates how mass measurements of massive stars in combination with the hadronic baseline assumed would further constrain $\ntrans$ (and simultaneously the strength of transition $\De\ep/\etrans$). 

At fixed stiffness $\cQMsq$ ($=0.6$ as an example in Fig.~\ref{fig:Mt_Dt_Mmax}) in high-density quark matter, a stiff hadronic EoS such as DBHF gives rise to heavier stars, and therefore allows a wide range of onset mass $\Mtrans$ and strength of transition $\De\ep/\etrans$ to be compatible with the $\Mmax\geq2\,\Msolar$ constraint (blue solid contour with triangles). 
For the softer HLPS EoS, a much larger region of the parameter space is eliminated, and for some densities there is even no valid transition with finite $\De\ep>0$ (red solid contour with diamonds). In such a scenario, a phase transition at too-high density, $\gtrsim 4\,n_0$, (the part of contour to the right) always leads to a short hybrid star branch connected to the normal hadronic star branch on the $M$-$R$ diagram and thus barely affects the measured quantities. 
This feature is unavoidable despite the fact that it maintains compatibility with observational data as long as the hadronic part is well-suited; see e.g. Ref.~\cite{Annala:2019puf}. We therefore disregard this high-density range in which the quark cores are masqueraded~\cite{Alford:2004pf}, and only consider the part of the contour to the left ($\ntrans\lesssim4\,n_0$) where the resulting hybrid stars are fairly distinctive. 
Dotted lines in this figure indicate how the excluded region would grow if a $\sim2.3\,\Msolar$ neutron star were to be observed. 
For HLPS the transition mass is pushed to exceedingly low values,  $<0.5\,\Msolar$. This suggests that all NSs that have been detected are actually hybrid stars. The mass-radius curves of such configurations are shown in Fig.~\ref{fig:MR_Mmax23} (a) of Appendix~\ref{AppB}. 
There high-mass stars have larger radii than canonical-mass stars, in contradiction with the standard assumption that $R$ remains roughly constant from $1-1.8\,\Msolar$ until close to the maximum-mass with a decrease in $R$ by $\sim 1$~km.

A similar significant reduction of usable parameter space for phase transition can happen if quark matter is assumed soft (with smaller $\cQMsq$), which results in a more limited range of $\ntrans$. For instance, the green dashed lines in Fig.~\ref{fig:nmax_ntrans} (b) (DBHF $\to \cQMsq = 0.4$) are shorter than the blue dashed lines (DBHF $\to \cQMsq = 0.6$). 
With softer quark matter, the transition density $\ntrans$ has to be small even if the low-density hadronic part is stiff, and increasing the lower bound on $\Mmax$ intensifies this requirement. Such stiff hadronic matter $\to$ soft quark matter transition usually predicts larger radii than normal hadronic stars at high masses, see e.g. Fig.~\ref{fig:MR_Mmax23} (b) and previous discussion on Figs.~\ref{fig:R20_base} and~\ref{fig:R20_more}.

It is worth emphasizing that extrapolating the EoS from nuclear regimes to densities in neutron stars is nontrivial. Cores of the lightest NSs ($\sim 1.1 \,\Msolar$) observed probably probe densities $\gtrsim 2.0 \,n_{0}$ (see Table~\ref{tab:had_EoS}), where uncertainties concerning nucleon-nucleon interactions already become considerably large.
If a strong first-order transition occurs below $2.0\, n_0$, then (subject to the maximum-mass star supported) the inferred radii for both heavy pulsars and canonical-mass stars are highly sensitive to the stiffness in quark matter. In some cases $\Rtwo\geq\Rtyp$ can be achieved. 
Ref.~\cite{Zhao:2020dvu} recently pointed out that in quarkyonic stars with hadron-quark crossover this is also a viable possibility, although for a different reason. 
A prominent stiffening ($c_s^2\gtrsim 0.8$) in the continuous EoS around $2.0 \,n_0$ where quarkyonic matter emerges results in masses and radii (with central densities $\gtrsim 2.0 \, n_{0}$) that are both larger compared to those without quarks. 
In contrast, hybrid stars with a first-order transition at sufficiently low densities are more similar to self-bound quark stars with a thin nuclear mantle. 
Nevertheless, combined information on $\Rtwo$ and $\Rtyp$ from observations will be vital to establish further constraints.

For EoSs with continuous energy density, it is likely that the smallest $\Rtwo$ that can support a $2\,\Msolar$ star is $\geq 9.5$~km with the minimum value determined by $\Mmax=2\,\Msolar$ (see e.g. Fig. 1 in Ref.~\cite{Annala:2017llu}). With the mass measurement of PSR J0740+6620 $\simeq 2.14\,\Msolar$, this lower bound is expected to increase. 
It would also be interesting to see how the overall relation between $\Rtwo$ and $\Mmax$ investigated in the present study would be modified if a broader template for hadronic EoSs were to be utilized, especially when a phase transition takes place at the central densities of canonical-mass stars that lie above the well-constrained nuclear regime. 
The minimum $\Rtwo$ of hybrid stars that are obtained specifically with a strong sharp transition at saturation $\ntrans=n_0$, on the other hand, turn out to be least affected by the hadronic matter baseline assumed and thus can be viewed as robust.

\section{Prospects for Future Constraints}
\label{sec:obs}

Future GW events of inspiraling binary neutron stars will update estimates on NS tidal parameters which then translate into constraints on NS radii assuming appropriate EoS priors~\cite{Abbott:2018exr,Abbott:2018wiz}. 
GW detections outperform x-ray observations in inferring radii owing to smaller systematic uncertainties in the determination of tidal properties. Depending on the chirp masses of the detected systems, information on the radii of canonical-mass stars $R_{1.1-1.7}$ or high-mass stars $R_{\gtrsim1.7}$ can be extracted, although for more massive stars it is less constraining due to the intrinsic smaller deformations 
(tidal deformability $\La\propto k_{2}(R/M)^{5}$ where the tidal Love number $k_2$ and the compactness $R/M$ both decrease steadily for massive stars~\cite{Hinderer:2009ca,Postnikov:2010yn})
in more compact binaries that challenge the sensitivity of detectors~\cite{Abbott:2020uma}. 
With more complete EoS representations that properly incorporate strong phase transitions, multiple events of neutron star binaries are capable of limiting the strength and onset density of the transition if indeed realized in binary components of the detected population~\cite{Chatziioannou:2019yko}. 
A single loud event such as GW170817 can neither identify nor rule out phase transitions despite being informative to constrain EoSs of hadronic matter only. For example, stiff hadronic EoSs with large radii incompatible with GW170817 can be revived if a strong phase transition softens the EoS and decreases radii in the relevant mass range of component stars \cite{Han:2018mtj}. 

A neutron-star-black-hole (NSBH) coalescence can also be promising. If the black hole is small and 
has a large spin, there is hope to measure the neutron star tidal parameters~\citep{Lackey:2013axa}. However, in highly asymmetric systems (e.g. GW190814~\citep{Abbott:2020khf} but only if its secondary component is indeed a neutron star), the neutron star undergoes minimal disruption before plunging into the black hole, and therefore finite-size effects are hardly measurable~\citep{Lackey:2013axa}.  
 
Finally, potential future measurements of post-merger GW signals will help unveil the dynamic evolution of merger remnants which is much more sensitive to the composition of matter, and can be combined with concurrent EM emission or pre-merger GW information. 

Heavy pulsars with masses $\gtrsim 2\,\Msolar$ play a central role in revealing EoS properties at very high densities $\gtrsim 4-8\, n_{0}$. If there are no abrupt density discontinuities in stellar matter, such masses imply high sound speed at high densities, presumably $c_s^2\gtrsim 0.7$ (see e.g. Refs.~\cite{Tews:2018kmu,Fujimoto:2019hxv}), and larger radii for the maximum-mass stars. 
An even higher lower bound on $\Mmax$ further strengthens the limitation on the sound speed. Given the small tidal deformability in GW170817 that favors soft matter at intermediate densities, the required rapid stiffening in the EoS to reach $\geq2.6\,\Msolar$ risks breaking the causal limit thus disfavoring the ``NSBH'' scenario for GW190814~\cite{Abbott:2020khf}.
If there exist finite discontinuities in the energy density such as in self-bound stars or hybrid stars with a sharp interface between bulk hadronic and quark phases, the tight constraint on $c_s^2$ is relaxed, and even with not-too-stiff quark matter ($\cQMsq\gtrsim0.4$) massive stars $\geq2\,\Msolar$ can be readily obtained with the transition density $\ntrans\lesssim 3.0\,n_0$ likely within reach of the interior of canonical-mass stars \cite{Han:2019bub}. 
As indicated by the results in Fig.~\ref{fig:R20_base} (a) and Fig.~\ref{fig:R20_more} (a), $\Mmax\geq 2.6\,\Msolar$ is still within the reach for a plausible range of parameters. Moreover, a small sound speed in quark matter (soft quark EoS) indicates large stars, opposite to that found for continuous EoSs. Consequently, a feasible lower bound on $R_{\sim 2.0}$ from e.g., {\it{NICER}} data would have different interpretations for these different scenarios of EoSs.

Using empirical relations established in hydrodynamical simulations that relate the radius $R$, the maximum mass $\Mmax$, and the threshold binary mass $M_{\rm thres}$ for prompt collapse, Ref.~\cite{Bauswein:2017vtn} derived independent lower limits on $R_{1.6}>10.68$~km and $\Rmax>8.6$~km, although the results rely on models in which strong phase transitions are not accounted for. 
It is expected that an upper limit on $\Rmax$ can also be extracted if high-mass mergers (e.g., GW190425 with total mass $\sim3.4\,\Msolar$~\cite{Abbott:2020uma}) are assumed to result in a prompt collapse (see their Fig. 3), which would tighten the limit on $R_{\sim 2.0}$ that is complementary to future {\it{NICER}} measurements of massive pulsars. Inspired by this possibility, we show supplementary plots in Appendix~\ref{AppA} that relate $\Rmax$ and $\Mmax$ for our calculations with hybrid EoSs, similar to previous plots showing $\Rtwo$ and $\Mmax$.

Other static observables that have the potential to determine NS radius include 
(1) a future measurement of moment of inertia in the double pulsar system PSR J0737-3039~\cite{Lattimer:2004nj,Landry:2018jyg}, 
(2) NS binding energy that can be inferred from a core-collapse supernova explosion through neutrino detection~\cite{Lattimer:2000nx}, and
(3) electron-capture supernova formation scenario for light neutron stars in binaries~\cite{Podsiadlowski:2005ig}.

\section{Conclusions}
\label{sec:con}

{\it{NICER}} measurements of the radii of $\sim 2\,\Msolar$ stars are promising to constrain hybrid EoSs with a sharp first-order hadron-quark transition, primarily the stiffness in quark matter and the critical mass above which neutron stars harbor a quark core. 
Our main results are contained in Figs.~\ref{fig:R20_base} and \ref{fig:R20_more}. 
The minimum values of $\Rtwo$ (and also $\Rmax$) are determined by the lowest possible transition density at saturation $\ntrans=n_{0}$, and these values will be larger if the lower bound on the maximum-mass $\Mmax$ increases. 
We find that a lower bound on $\Mmax$ with an upper bound on $\Rtwo$ eliminates too soft quark matter, and an upper bound on $\Mmax$ with a lower bound on $\Rtwo$ strongly disfavors a phase transition into too-stiff quark matter appearing at low densities.

Guided by laboratory experiments and nuclear theory, the behaviors of hadronic EoSs assumed up to $\sim 2\,n_0$ have more impact on the canonical-mass stars (for which various astrophysical and gravitational-wave constraints are available) than for very massive stars. Many of the constraints derived thus far have not taken into account possible phase transitions. 
The CSS parametrization for sharp phase transitions can be combined with other representations of the hadronic EoSs, such as the spectral decomposition~\cite{Lindblom:2012zi,Lindblom:2013kra}, piece-wise polytropic extrapolation~\cite{Read:2008iy, Hebeler:2013nza}, modeling of sound speed with Gaussian functions~\cite{Greif:2018njt,Tews:2018kmu}, and an expansion of symmetry energy terms at high densities~\cite{Xie:2020tdo}, etc. toward a more complete analysis in the future.

We find that combining constraints on $\Rtyp$ and $\Rtwo$ lead to different interpretations in different scenarios of the EoS. 
If there is evidence for $\Rtwo>\Rtyp$, it likely suggests either a high sound speed ($c_s^2\gtrsim 0.7$) in the EoS for which energy density is continuous, as in standard hadronic matter or in quarkyonic matter with a hadron-quark crossover, or a strong first-order phase transition has taken place below the central density of $\lesssim 1.1 \,\Msolar$ stars, i.e., all NSs are hybrid stars (see e.g. Fig.~\ref{fig:MR_Mmax23}). 
Note that self-bound strange quark stars naturally fulfill this condition as well, but they cannot account for the surface emission observed from NSs. 
With sufficiently stiff quark matter hybrid stars that exhibit a strong transition tend to be more compact than their hadronic counterparts, and the difference between $\Rtwo$ and $\Rtyp$ can be larger than normally achieved in hadronic models (see e.g. Fig.~\ref{fig:MR_Mmax20} (a)). 

We wish to emphasize the importance of combining future $R_{\sim 2.0}$ constraint, heavy pulsar mass measurements that limit $\Mmax$, and available radius estimates of $\sim 1.4\,\Msolar$ stars from different sources/messengers including GW and x-ray observations to eventually narrow down the variations of theoretical models to explain them. 
In the hadronic matter only scenario, mass-radius constraints from J0030+0451 by {\it{NICER}}~\cite{Riley:2019yda,Miller:2019cac} favor a stiffer EoS and the binary tidal deformability constraint from GW170817 \cite{Abbott:2018wiz,Abbott:2018exr} favors a softer EoS. 
A joint analysis of data from both improves our knowledge on the overlapping constraints that are consistent~\cite{Landry:2020vaw,Essick:2020flb,Jiang:2019rcw}.  
An alternative explanation would be that one or both of the component stars of GW170817 are hybrid stars with a strong phase transition inducing smaller tidal deformation, whereas J0030+0451 is a normal hadronic star with larger radius~\cite{Christian:2019qer}. 
By incorporating EoSs with strong phase transitions, a preference for the standard or alternative explanation can be quantitatively evaluated, although GW170817 alone is not yet informative enough~\cite{Essick:2020flb}. Further constraints on $R_{\sim 2.0}$ and $\Mmax$ would help distinguish between these different scenarios. Numerical simulations of binary mergers and their EM counterpart modelings will also provide support to reveal additional information on the merger remnant, such as $\Mmax$ and $\Rmax$.     


\section*{Acknowledgements}

We thank Katerina Chatziioannou for useful comments and suggestions. 
S.H. is supported by the National Science Foundation, Grant PHY-1630782, and the Heising-Simons Foundation, Grant 2017-228. M.P. is supported by the Department of Energy, Grant No. DE-FG02-93ER-40756. 
S.H. would like to thank the participants of the Second Nuclear and Particle Theory Meeting held virtually at Washington University in St. Louis for illuminating discussions.
%

\appendix
\section{Relating $\Rmax$ and $\Mmax$}
\label{AppA}

\begin{figure*}[]
\parbox{0.48\hsize}{
\includegraphics[width=\hsize]{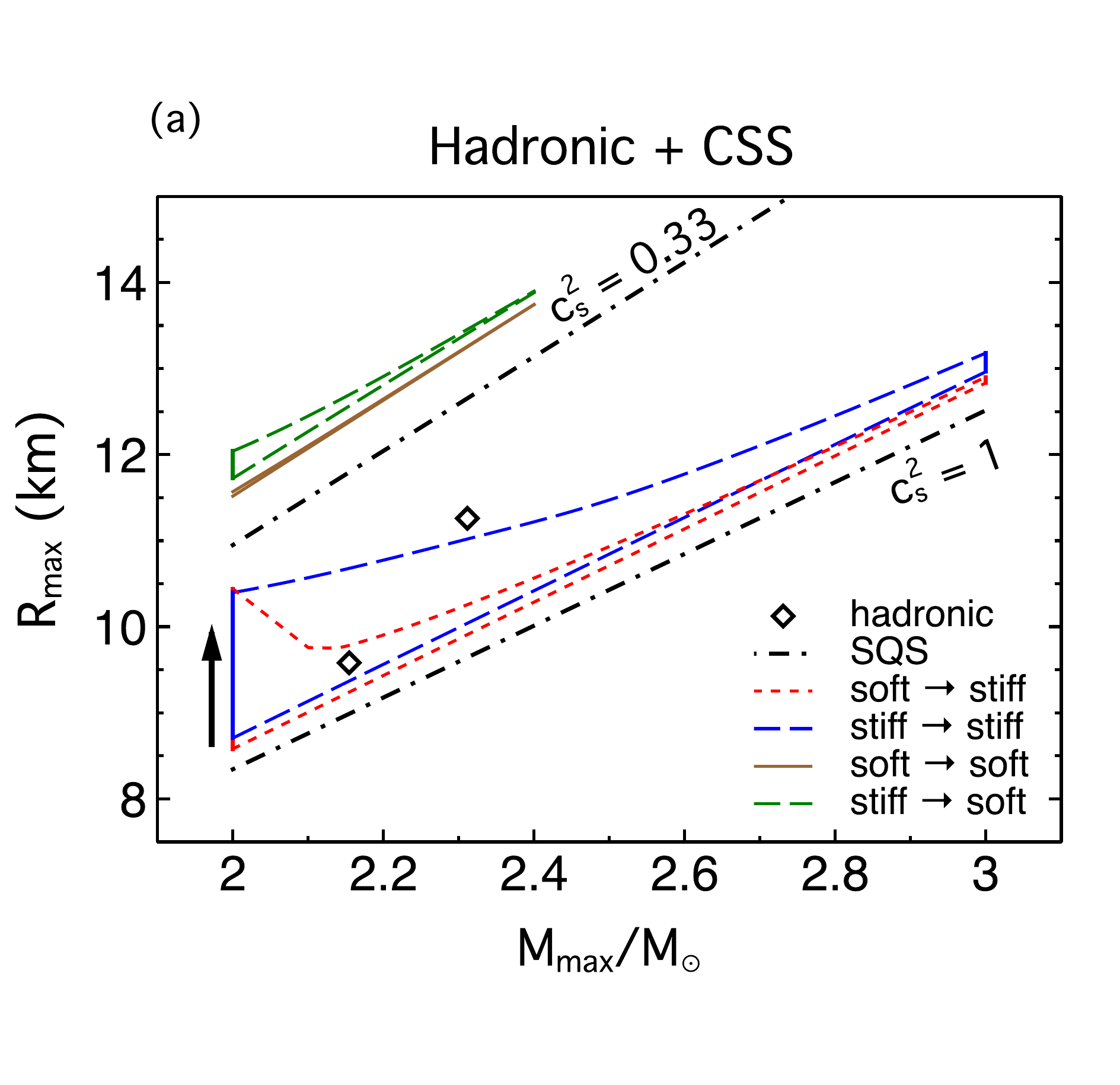}\\[-2ex]
}\parbox{0.48\hsize}{
\includegraphics[width=\hsize]{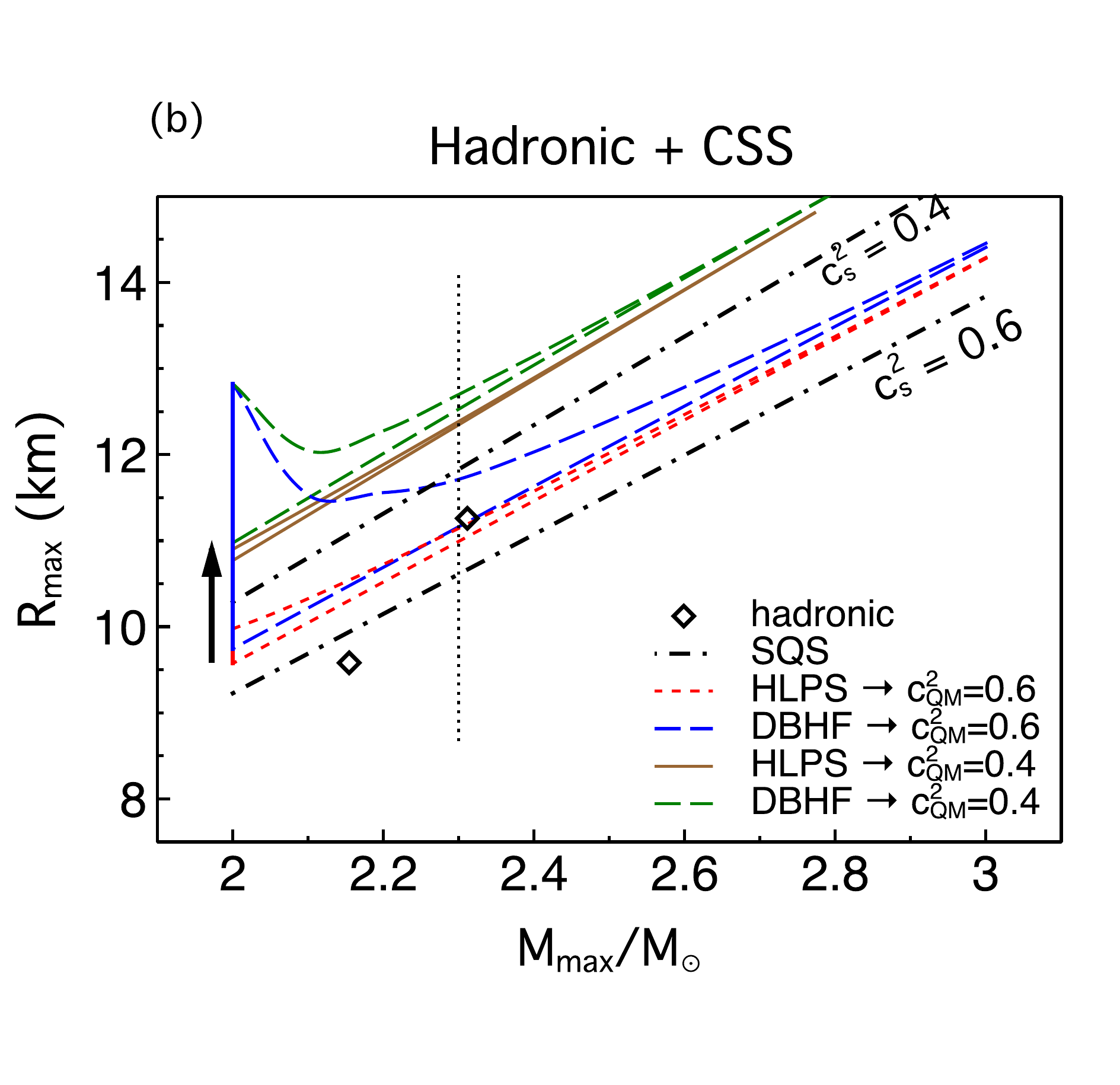}\\[-2ex]
}
\caption{Similar to the left panels of Figs.~\ref{fig:R20_base} and \ref{fig:R20_more}, but with $y$-axis showing $\Rmax$, the radius of the heaviest star.
}
\label{fig:Rmax_Mmax}
\end{figure*}

As in Figs.~\ref{fig:R20_base} (a) and \ref{fig:R20_more} (a), Fig.~\ref{fig:Rmax_Mmax} shows the radius $\Rmax$ vs. the mass $\Mmax$ of the heaviest star, but with the same hybrid EoSs used. 
The upward arrows indicate the direction along which the transition density $\ntrans$ (and thus $\Mtrans$) increases. 
In panel (a), the leftmost boundaries of the four colored bands include all configurations with $\Mmax=2.0\,\Msolar$. In panel (b), the vertical dotted line marks $\Mmax=2.3\,\Msolar$. For both cases, typical $M$-$R$ plots are provided in Appendix~\ref{AppB}.

\section{Mass-radius plots for typical EoSs}
\label{AppB}

\begin{figure*}[]
\parbox{0.48\hsize}{
\includegraphics[width=\hsize]{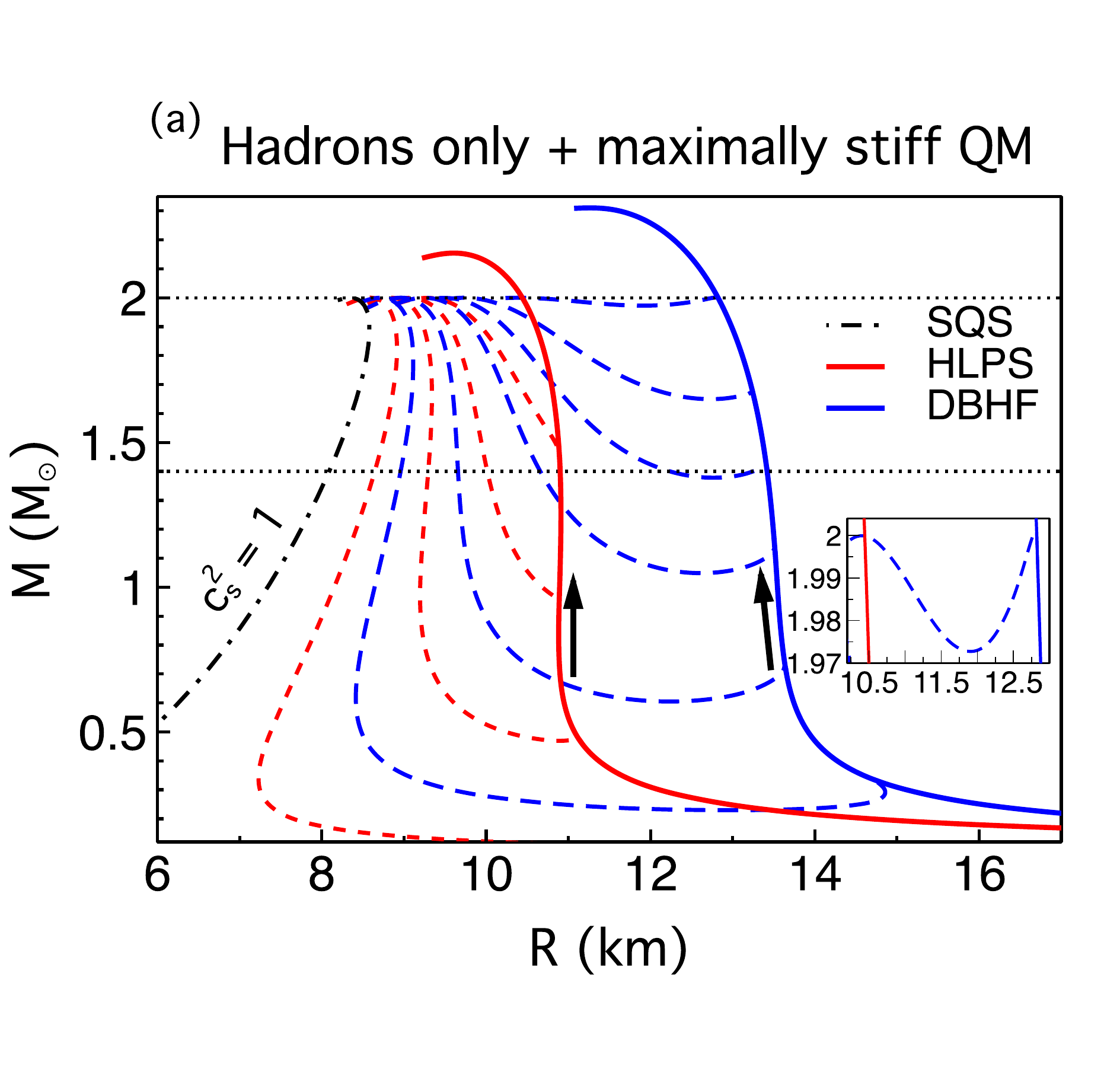}\\[-2ex]
}\parbox{0.48\hsize}{
\includegraphics[width=\hsize]{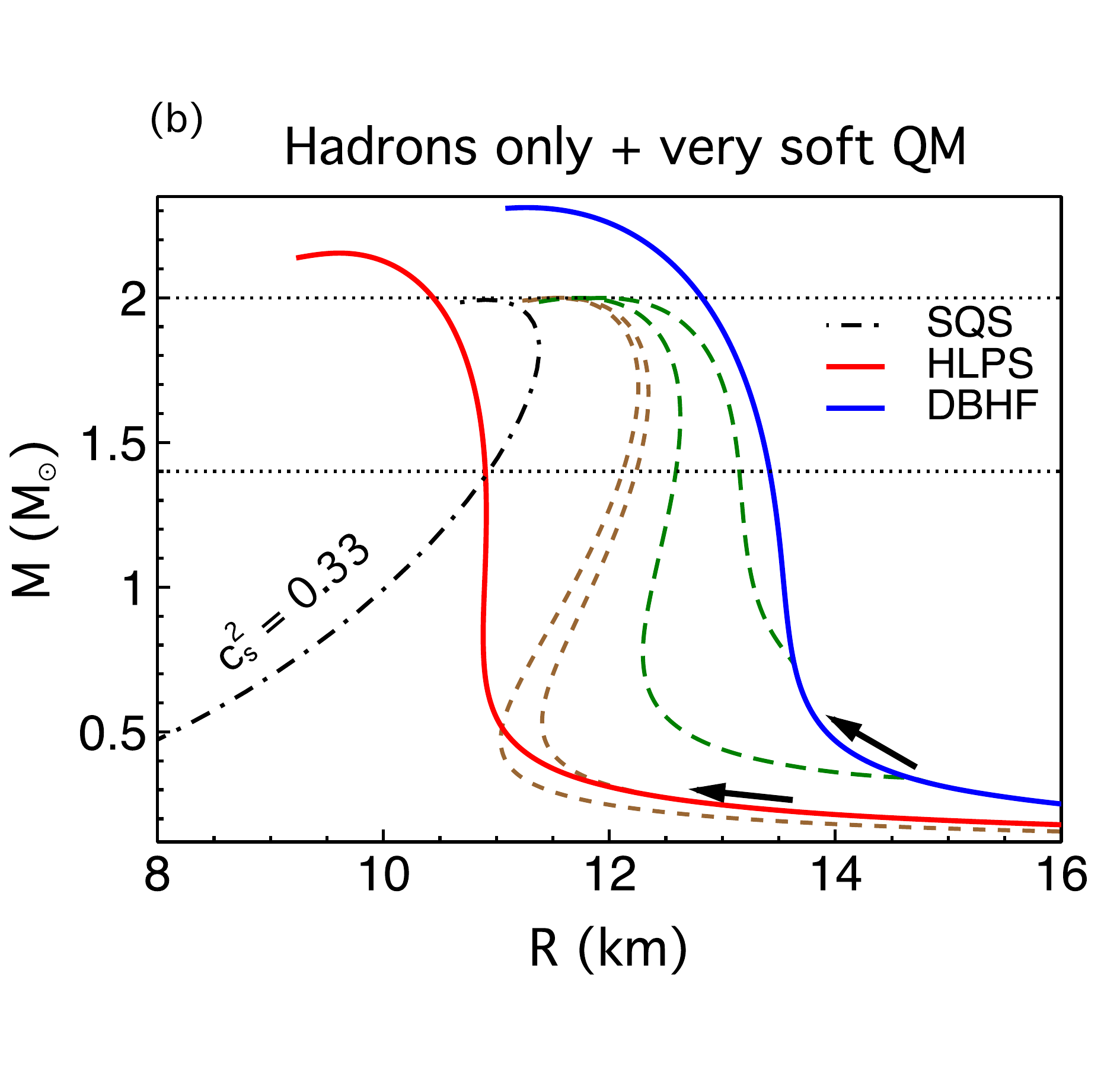}\\[-2ex]
}
\caption{Left panel: $M$-$R$ curves with transition densities $\ntrans/n_{0}=1.0,~2.2, ~3.0,$ and $3.9$ for HLPS (soft) coupled with  maximally stiff quark matter ($\cQMsq=1$); for DBHF (stiff) and $\cQMsq=1$, $\ntrans/n_{0}=1.0, ~1.5, ~2.0, ~2.4, ~2.8,$ and $~3.536$. 
Right panel: here the transition densities are $\ntrans/n_{0}=1.0$ and $1.5$ for HLPS (soft) in combination with very soft quark matter ($\cQMsq=0.33$); for DBHF (stiff) with the same $\cQMsq$, $\ntrans/n_{0}=1.0$ and $1.5$. 
In both cases, the maximum mass is fixed at $\Mmax=2.0\,\Msolar$ as in the leftmost boundaries of colored bands in Fig.~\ref{fig:R20_base} (a). 
Note that in the left panel, there exists a special twin-star configuration (inset plot) for stiff DBHF $\to$ stiff quark matter with both the heaviest hadronic star and the heaviest hybrid star having the same $M=2.0\,\Msolar$. With maximally stiff quark matter, a majority of hybrid EoSs leads to ``third-family'' stars as illustrated in Fig.~\ref{fig:MR-illust} (c).
}
\label{fig:MR_Mmax20}
\end{figure*}

\begin{figure*}[]
\parbox{0.48\hsize}{
\includegraphics[width=\hsize]{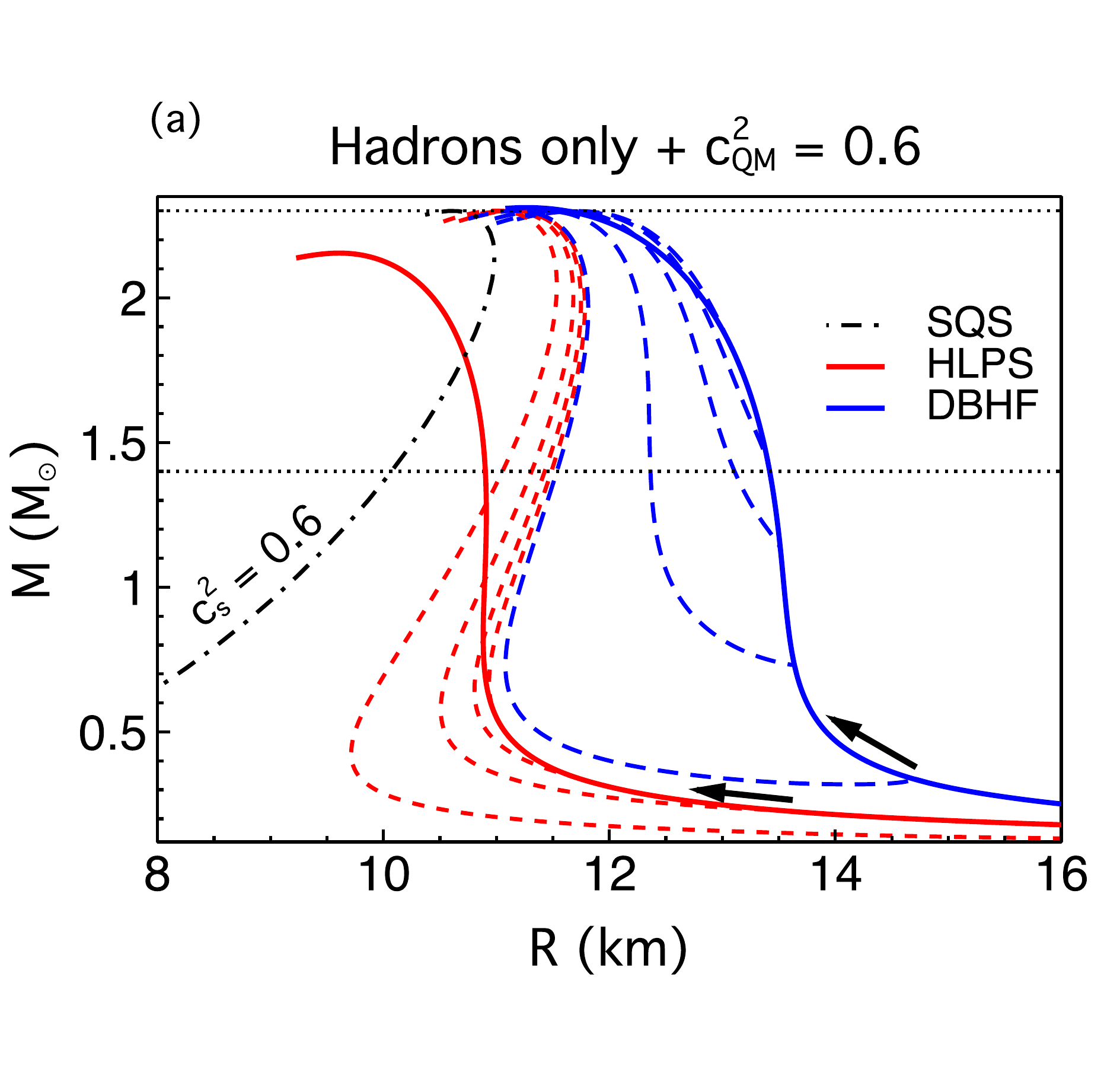}\\[-2ex]
}\parbox{0.48\hsize}{
\includegraphics[width=\hsize]{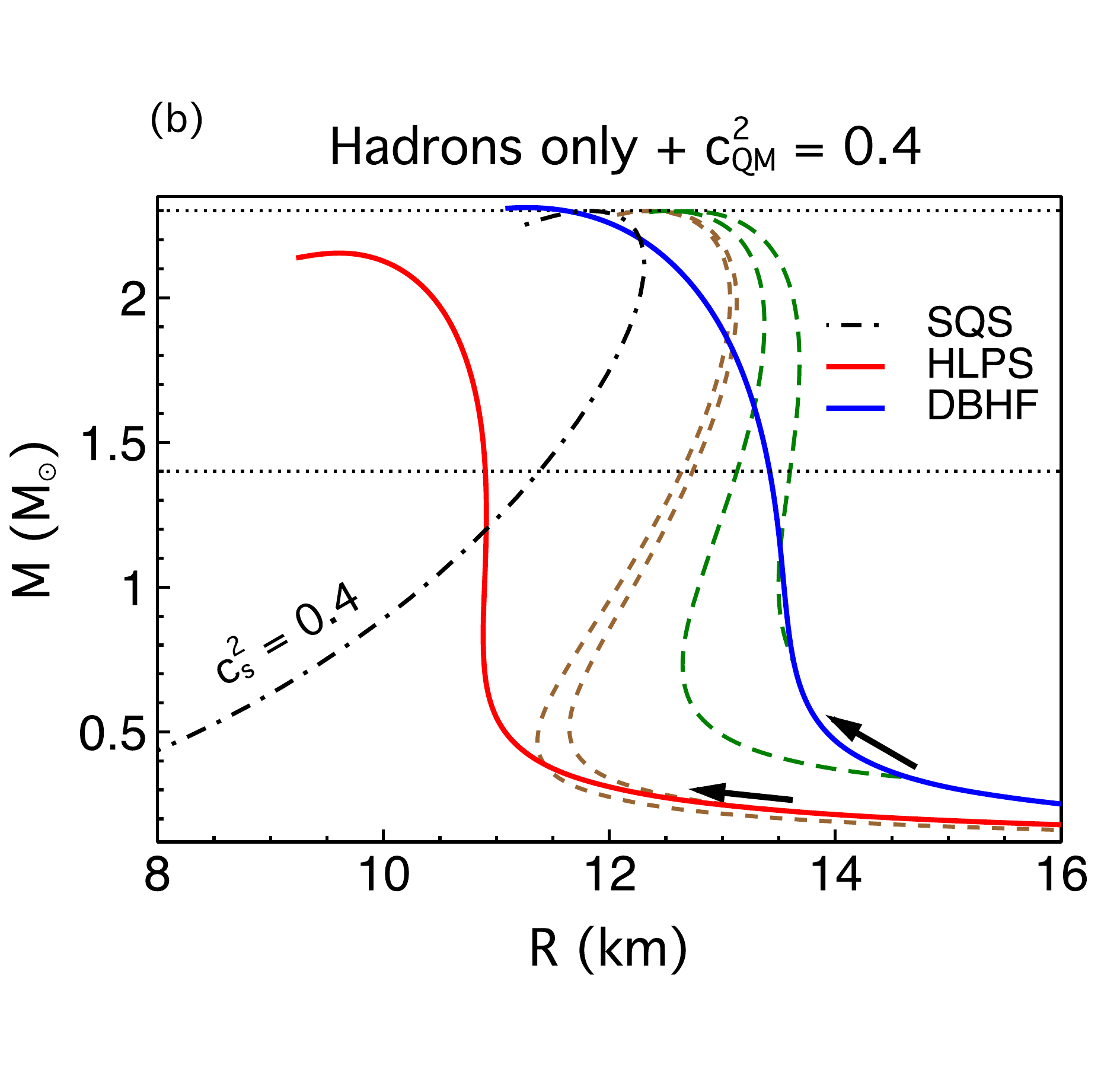}\\[-2ex]
}
\caption{Left panel: $M$-$R$ curves with transition densities $\ntrans/n_{0}=1.0,~1.6, ~1.9,$ and $~2.2$ for HLPS (soft) coupled with quark matter with $\cQMsq=0.6$; 
for DBHF (stiff), $\ntrans/n_{0}=1.0, ~1.5, ~2.0, ~2.4,$ and $~2.8$. 
Right panel: HLPS (soft) $\to$ quark matter with $\cQMsq=0.4$, $\ntrans/n_{0}=1.0$ and $1.476$; for DBHF (stiff), $\ntrans/n_{0}=1.0$ and $1.5$. 
In both cases, the maximum mass is fixed at $\Mmax=2.3\,\Msolar$, and is indicated by the dotted vertical line in Fig.~\ref{fig:Rmax_Mmax} (b). Mass-radius curves with $\Rtyp<\Rtwo$ (similar to case $(3)^{*}$ in Fig.~\ref{fig:MR-illust} (b)) refer to hybrid EoSs for which the transition takes place at low-enough densities and the high-density quark matter is not too stiff.
}
\label{fig:MR_Mmax23}
\end{figure*}

Figure~\ref{fig:MR_Mmax20} contains $M$-$R$ curves for representative EoSs from Fig.~\ref{fig:R20_base}. In all cases, the maximum mass of hybrid stars are fixed at $\Mmax=2.0\,\Msolar$, and the associated mass contours can be found in Fig.~\ref{fig:R20_base} (b) and Fig.~\ref{fig:nmax_ntrans} (a). Figure~\ref{fig:MR_Mmax23} shows similar plots but for selected EoSs from Fig.~\ref{fig:R20_more}, and with a different $\Mmax=2.3\,\Msolar$. The associated mass contours are in Fig.~\ref{fig:R20_more} (b) and Fig.~\ref{fig:nmax_ntrans} (b).

\bibliographystyle{apsrev4-1}
\bibliography{R20_v2}

\end{document}